\documentclass[aip,apl,unsortedaddress,superscriptaddress,reprint]{revtex4-1}
\usepackage{hyperref}
\usepackage{graphicx}
\usepackage{amsmath}
\usepackage{amssymb}
\usepackage{times,mathptmx}
\usepackage{dcolumn}
\usepackage{ifthen}
\usepackage{color}
\usepackage{proof}

\newcommand{\ie}{i.\,e.}

\newcommand{\degree}{\ensuremath{\circ}}

\newcommand{\thetaSH}{\ensuremath{\theta_\mathrm{SH}}}

\newcommand{\jsd}{\ensuremath{\mathbf{j}_\mathrm{s}}}

\newcommand{\lspin}{\ensuremath{\lambda_\mathrm{sf}}}
\newcommand{\lspinNM}{\ensuremath{\lambda_\mathrm{sf,NM}}}

\newcommand{\jqd}{\ensuremath{\mathbf{j}_\mathrm{q}}}

\newcommand{\jhd}{\ensuremath{\mathbf{j}_\mathrm{h}}}
\begin{document}

\title[]{All-electrical Magnon Transport Experiments in Magnetically Ordered Insulators}

\author{Matthias Althammer}
\email{matthias.althammer@wmi.badw.de}
\affiliation{Walther-Mei{\ss}ner-Institut, Bayerische Akademie der Wissenschaften, 85748 Garching, Germany}
\affiliation{Physik-Department, Technische Universit\"at M\"unchen, 85748 Garching, Germany}

\date{\today}
\begin{abstract}
Angular momentum transport is one of the cornerstones of spintronics. Spin angular momentum is not only transported by mobile charge carriers, but also by the quantized excitations of the magnetic lattice in magnetically ordered systems. In this regard, magnetically ordered insulators provide a platform for magnon spin transport experiments without additional contributions from spin currents carried by mobile electrons. In combination with charge-to-spin current conversion processes in conductors with finite spin-orbit coupling it is possible to realize all-electrical magnon transport schemes in thin film heterostructures. This review provides an insight into such experiments and recent breakthroughs achieved. Special attention is given to charge current based manipulation via an adjacent normal metal of magnon transport in magnetically ordered insulators in terms of spin-transfer torque. Moreover, the influence of two magnon modes with opposite spin in antiferromagnetic insulators on all-electrical magnon transport experiments is discussed.
\end{abstract} 
\maketitle


\section{Introduction}
In solid state physics, electrons are conduits that not only transport charge and heat, but also spin, giving rise to many fascinating transport phenomena. In case of electronic transport charge and spin are inherently connected to each other and provide a basis to investigate spin transport by charge transport experiments. This dates back to the late 19th century with the discovery of the anisotropic magnetoresistance~\cite{Thomson1857,McGuire1975} and the anomalous Hall effect~\cite{Hall1881,nagaosa_anomalous_2010} in ferromagnetic 3d transition metals. A full quantitative microscopic description of both phenomena represents still a challenge today, but these two experimental observations were the spark to ignite the field of spintronics~\cite{Zutic2004}. Spintronic exploits the spin degree of freedom for information processing and storage. With the advances in thin film technology in the 1960s and 1970s first integration concepts and applications were realized, but ultimately limited in sensitivity by the small resistance changes of at those times known magnetoresistance effects. With the discovery of the giant magnetoresistance effect~\cite{Binasch1989,Baibich1988,Parkin1995} and the tunneling magnetoresistance~\cite{Julliere1975,Moodera1995,Miyazaki1995,Parkin2004,Yuasa2004} the field of spintronics had its first big success stories. These new magnetoresistance effects significantly enhanced sensitivity in magnetic field sensors and provided the means to keep pace with the exponentially increasing demands for data storage capacity associated with our transition into the age of information technology.

In these first spintronic applications, everything revolved around spin-polarized charge transport carried by conduction electrons. This concept inherently suffers from the same problems as conventional electronics, predominantly an increase in Joule heating and power consumption, when reducing the structure size of the devices. Yet, the demand for higher information processing speeds, denser information storage and low power consumption in information technology is ever increasing in our information driven society, such that exploring alternatives to conventional electronics and concepts beyond the von Neumann architecture is now in the spotlight of research~\cite{bourianoff_nanoelectronics_2010,waldrop_chips_2016}. Within the realm of magnetism and spintronics, the last two decades delivered a multitude of new concepts spanning from faster spin-based information processing, over efficient spin-manipulation schemes, to neuromorphic computing schemes \cite{sander_2017_2017,coll_towards_2019,Grollier2020,Vedmedenko2020}. Among these novel concepts are pure spin currents, representing the flow of spin angular-momentum without an accompanying charge current~\cite{hoffmann_pure_2007,Bauer2012,althammer_2018,Shen2019}.

The concept of pure spin currents relies on the fact that spin angular momentum is not only carried by electrons but also other particles and quasi-particles in solid state systems. Especially in magnetically ordered systems the quantized, bosonic excitations of the magnetic lattice, \ie~magnons, can be used to transport spin information across long distances even in electrical insulators. Owing to their long magnon lifetime, magnetically ordered insulators (MOIs) represent the basis for spin information processing via magnons (spin waves) in wave-based approaches forming the field of magnonics~\cite{Chumak2015,Rezende2020,Brataas2020}. Interestingly, experiments confirmed that the inherent Bose-Einstein statistics of magnons can be exploited to form magnon Bose-Einstein condensates providing access to magnon supercurrents and new effects inspired by superconductivity~\cite{Demokritov2006,Demidov2007,nakata_spin_2017,Bozhko2019}. Moreover, pure spin currents carried by magnons are not only possible in systems with ferromagnetic order, but also in antiferromagnetic systems~\cite{Rezende2016,Shen2019}. Due to the two magnetic sublattices of antiferromagnets, magnons come in pairs with opposite spins, such that antiferromagnetic insulators provide access to rich electronics inspired phenomena. In addition, even electrical conductors without magnetic order, but with finite spin-orbit coupling, so called normal metals (NM), can be employed to generate and detect pure spin currents via electrical charge currents. Among the effects for this charge-to-spin current conversion, the well known spin Hall effect~\cite{Dyakonov1971,Hirsch1999} has been employed to reliably switch magnetization in ferromagnets or even drive auto-oscillations in magnetically ordered materials via an adjacent conductor with large spin-orbit coupling~\cite{Miron2011,Demidov2012,Avci2016,Cheng2016,chen_spin-torque_2016,safranski_spin_2017}. Moreover, the spin Hall effect enables the electrical sensing of the orientation of the magnetic order parameter via the spin Hall magnetoresistance in NM/MOI heterostructures~\cite{Nakayama2013,althammer_quantitative_2013,chen_theory_2013,Hahn2013SMR,Chen2016SMRReview,althammer_2018}. Last but not least, topology provides an intriguing perspective on pure spin currents ranging from the quantum spin Hall effect in topological insulators~\cite{Bernevig2006,Chang2013,cazalilla_quantum_2014,qian_quantum_2014,Chang2015,Feng2015,Liu2016,Tokura2019} to pure spin current transport via skyrmions in topological magnetic textures~\cite{Muhlbauer2009,Neubauer2009,jiang_blowing_2015,Fert2017,EverschorSitte2018,Topology2018,Zzvorka2019,Back2020}. As evident from this short overview, pure spin current physics are covered in a variety of areas in magnetism and spintronics and tremendous progress has been achieved in this field over the last years. Thus, pure spin currents carry the promise for even more novel phenomena to be envisioned and realized in the future and represent an exciting field to work on as a researcher.

In this review we discuss the physical principles of all-electrical magnon transport experiments in magnetically ordered insulators. In the following, we first describe the mechanisms exploited to drive and detect magnon transport in magnetically ordered insulators by applying a charge current and measuring a voltage in adjacent electrical conductor strips via spin-to-charge conversion processes and interfacial spin currents. As a next step, we provide more details on the diffusive magnon transport process in the MOI. This is followed up by an introduction into how charge currents also allow for a manipulation of the magnon transport and how such changes are detected in the experiment. Next, we discuss how in antiferromagnetic insulators coupling between the two magnon modes with opposite spin lead to the manifestation of the magnon Hanle effect. Finally, we provide an outlook into future directions and opportunities for all-electrical magnon transport.
\section{Electrically-driven magnon transport in magnetically ordered insulators}
As previously discussed not only electrons allow to transport spin angular momentum, but also excitations of the magnetic lattice (magnons) can transport spin currents over $\mathrm{\mu m}$ long distances. Based on the theoretical prediction by Zhang and Zhang~\cite{ZhangMMR1,ZhangMMR2} and the pioneering experimental work by Cornelissen \textit{et al.}~\cite{Cornelissen2015} it is possible to utilize electrical charge currents in a NM to investigate the transport of pure spin currents via magnons in MOIs, which we refer to as all-electrical magnon transport experiments. Several groups have already made tremendous contributions towards a better understanding of all-electrical magnon transport experiments in MOIs~\cite{Goennenwein2015,Cornelissen2016_temperaturedependence,Cornelissen2016_fielddependence,shan_influence_2016,Cornelissen2016,Das2017,Cornelissen2018,Klaui2018,Liu2018,Thiery_2018,Shen2019,wimmer_AHE_2019,Avci2020,Troncoso2020,Klaui2020,Han2020,Lebrun2020,Ross2020}.  These experiments utilize a heterostructure consisting of two NM strips in contact with the MOI as illustrated in Fig.~\ref{figure:MagnonTransport}. In the experiment a charge current density $\mathbf{j}_\mathrm{q}$ is applied to the NM injector strip, which generates a spin current density $\jsd$. At the interface the electron spin current leads to the injection of a diffusive magnon spin current into the MOI. In the MOI this diffusive spin current is transported towards the NM detector strip and detected as an open circuit voltage via a spin-to-charge current conversion process. In the following we will discuss in more detail these processes of electrical injection and detection and diffusive magnon transport in the MOI.

\subsection{Electrical injection and detection of magnons}
For the all-electrical magnon transport experiments we rely on effects enabling the generation and detection of pure spin currents $\mathbf{j}_\mathrm{s}$ via charge currents $\mathbf{j}_\mathrm{q}$ in electrical conductors with finite spin-orbit coupling. A very prominent example of such an effect in NMs is the spin Hall effect (SHE)~\cite{Dyakonov1971,Hirsch1999,Hoffmann2013,Sinova2015}. The SHE critically depends on spin-dependent scattering and bandstructure effects, such that electrons acquire a spin-dependent transverse velocity, when traversing through an electrical conductor with spin-orbit coupling. The transformation from $\mathbf{j}_\mathrm{q}$ to $\mathbf{j}_\mathrm{s}$ (with spin polarization $\mathbf{s}$) is described by~\cite{Hirsch1999,Hoffmann2013,Sinova2015}
\begin{equation}\label{eq:SHE}
\jsd=\frac{\hbar}{2 e} \thetaSH \jqd\times\mathbf{s}\;,
\end{equation}
where $\thetaSH$ describes the efficiency of the spin-to-charge current conversion process. The inverse process, called the inverse SHE (ISHE), a pure spin current $\jsd$ with spin polarization $\mathbf{s}$ is transformed into a charge current:
\begin{equation}\label{eq:iSHE}
\jqd=\frac{2e}{\hbar} \thetaSH \mathbf{s}\times\jsd\;.
\end{equation}
In the SHE we find an orthogonal relation between $\jqd$, $\jsd$, and $\mathbf{s}$. In a finite sized NM under application of $\jqd$ an electron spin accumulation (described via the electron spin-dependent chemical potential and on the lengthscale of the electron spin diffusion length) with spin orientation $\mathbf{s}$ will form at the surfaces of the NM, where the spin orientation, surface normal and charge current direction are orthogonal to each other. 

We want to note that we here only considered the case of normal metals, i.e.~conductors without a magnetic order. However, recent theoretical predictions and experiments have shown that also metals with magnetic order provide rich means to generate spin currents~\cite{Miao2013,Mendes2014,Zhang2014,Wu2015_AHE,Wang2014_AHE,Reichlov2015,Tian2016,ou_strong_2016,Das2017,Qin2017_AHE,Iihama2018,elezn2018,Baltz2018,Das2018,Amin2018,Omori2019_AHE,wimmer_AHE_2019,Kimata2019,Manchon2019,Wu2019,Das2020,Davidson2020,Mook2020,Santos2021}, that can interact with MOIs. Most strikingly, the orientation of the spin polarization is not necessarily along the direction of the magnetic order parameter, but can also be oriented perpendicular to it~\cite{Kimata2019,Davidson2020,Mook2020}. These transverse spins exist on the length-scale of the exchange length, quite similar to the spin-diffusion length in NMs with large spin-orbit coupling. Taken together, the utilization of ferromagnetic and antiferromagnetic conductors with spin-orbit coupling opens up new avenues to venture for pure spin current physics. In this regard, all-electrical magnon transport experiments already showed that they are beneficial to qualitatively and quantitatively investigate charge-to-spin current conversion processes in magnetically ordered conductors~\cite{Das2017,Das2018,wimmer_AHE_2019,Das2020,Santos2021}.

\begin{figure}
 \centering
 \includegraphics[width=85mm]{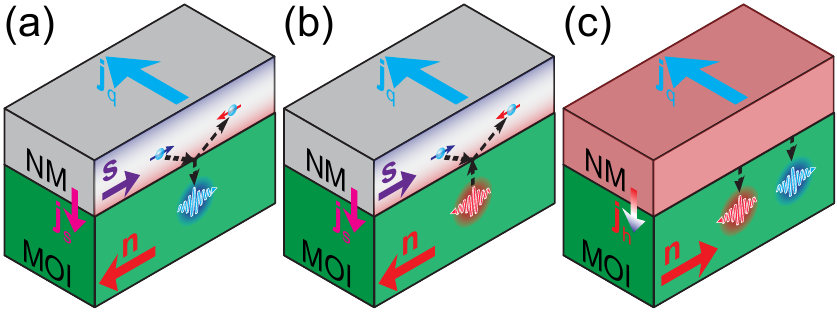}\\
 \caption[All-electrical magnon generation schemes]{Illustration of all-electrical magnon generation schemes in MOI/NM heterostructures. A charge current $\jqd$ is applied to the NM leading to a spin accumulation at the NM/MOI interface. (a) Generation of an $\alpha$-mode magnon via inelastic electron spin-flip scattering at the interface. (b) Inelastic electron spin-flip scattering in this configuration can also lead to the absorption of a $\beta$-mode magnon at the interface. (c) In addition, Joule heating via the applied charge current causes an injection of a heat current into the MOI (affecting the $\alpha$- and $\beta$-mode magnons simultaneously as carriers of heat) and in addition an injection of a pure spin current via the interfacial spin Seebeck effect.}
  \label{figure:MagnonGeneration}
\end{figure}

As a next step, we need to describe the transfer of spin momentum at the interface between the NM and the MOI. A detailed theoretical description is presented in~\cite{Bender2015} including also heat transport across the NM/MOI interface. In the following, we want to briefly discuss the consequences for interfacial spin transfer. We assume that within the NM a spin accumulation with spin orientation $\mathbf{s}$ is described via the spin-dependent chemical potential $\mu_\mathrm{s}(z)$ ($\mathbf{z}$ is oriented along the interface normal and $z=0$ at the interface). While in the MOI the magnetic order parameter $\mathbf{N}$ (and a unit vector describing its orientation $\mathbf{n}=\mathbf{N}/N$) describes the magnetic structure. In addition, a spin accumulation of magnetic excitation quanta is parameterized via the spin magnon chemical potential $\mu_\mathrm{mag,s}$. Moreover, we account for the temperature profile in the systems by assigning the temperatures $T_\mathrm{N}$ and $T_\mathrm{mag}$ to the electronic system in the NM and the magnonic system in the MOI, respectively. Utilizing these parameters, we can write for the amount and spin orientation of the pure spin current across the NM/MOI interface~\cite{Bender2015}
\begin{align}
\mathbf{j}_\mathrm{s,int}&=\frac{1}{4\pi}\left(\tilde{g}_i^{\uparrow\downarrow}+\tilde{g}_r^{\uparrow\downarrow}\mathbf{n}\times\right)\left(\mu_\mathrm{s}(0)\mathbf{s}\times\mathbf{n}-\hbar\dot{\mathbf{n}}\right) \nonumber \\ +&\left[g(\mu_\mathrm{mag,s}+\mu_\mathrm{s}(0)\mathbf{s}\cdot\mathbf{n})+S(T_\mathrm{mag}-T_\mathrm{N}) \right]\mathbf{n}\;.
\label{equ:InterfacialSpinCurrent}
\end{align}
Within this equation, the orientation of $\mathbf{j}_\mathrm{s,int}$ is the spin orientation of the spin current flowing across the interface and not the flow direction of the pure spin current, which is always oriented perpendicular to the interface along $\mathbf{z}$. Here, $\tilde{g}_i^{\uparrow\downarrow}$ and $\tilde{g}_r^{\uparrow\downarrow}$ are the imaginary and real parts of the effective spin mixing conductance, which already accounts for a finite temperature and the magnon bandstructure of the MOI. $g$ is the spin conductance and $S$ the spin Seebeck coefficient. All these four parameters can be calculated from the real and imaginary parts of the $T=0$ spin-mixing conductance $g^{\uparrow\downarrow}$~\cite{brataas_finite-element_2000,tserkovnyak_nonlocal_2005,jia_spin_2011,Bauer2012}, which describes the spin transparency of the interface, and taking into account the magnon density of states $D(E)$ given by the magnon bandstructure of the MOI~\cite{Bender2015,Cornelissen2016}. Only $\tilde{g}_i^{\uparrow\downarrow}$ and $\tilde{g}_r^{\uparrow\downarrow}$ remain finite at $T=0$, while $g$ and $S$ vanish. Thus, $g$ and $S$ are governed by thermal fluctuations. The physical principle behind the contributions to the spin current across the interface are elastic and inelastic spin-flip scattering processes at the interface for the mobile charge carriers in the NM. For  $\tilde{g}_i^{\uparrow\downarrow}$ and $\tilde{g}_r^{\uparrow\downarrow}$ elastic spin-flip scattering is the dominant contribution, while inelastic scattering only leads to additional corrections of $\tilde{g}_r^{\uparrow\downarrow}$~\cite{Bender2015,Cornelissen2016}. For the case of elastic spin-flip scattering, angular-momentum of the spin-flip is transferred via a torque $\mathbf{\tau}$ onto the magnetic order parameter $\mathbf{n}$. Both $g$ and $S$ originate from inelastic electron spin-flip scattering at the interface~\cite{Bender2015,Cornelissen2016} as illustrated in Fig.~\ref{figure:MagnonGeneration}. The associated change in angular momentum and energy of the charge carrier in the NM is transferred to magnetic excitation quanta in the MOI and thus couples $\mu_\mathrm{s}$ and $\mu_\mathrm{mag,s}$. Speaking in terms of spin-transfer torque, the spin accumulation in the NM now transfers angular momentum to thermal fluctuations of the magnetic lattice, \ie~it enhances or reduces the number of magnetic excitation quanta at the MOI/NM interface~\cite{Guo2018}. This process is maximized if $\mathbf{s}$ and $\mathbf{n}$ are collinear with each other. It is important to note that the spin orientation of the spin current across the MOI/NM interface caused by $g$ and $S$ is always oriented along the magnetic order parameter $\mathbf{n}$.

The spin current across the MOI/NM interface is crucial for the manifestation of the spin pumping effect~\cite{Tserkovnyak2002,tserkovnyak_nonlocal_2005,Costache_PRL2006,Mosendz2010,ando_inverse_2011,Heinrich:2011dk,Hahn2013}, the spin Seebeck effect~\cite{Uchida2008,Uchida2010,uchida_longitudinal_2013,Adachi2013,Uchida2014}, the spin Hall magnetoresistance~\cite{Nakayama2013,althammer_quantitative_2013,chen_theory_2013,Hahn2013SMR,Chen2016SMRReview} and all-electrical magnon transport experiments~\cite{Cornelissen2015,Goennenwein2015,Cornelissen2016}. In all experiments, the detection of these interfacial spin currents is realized by the fact that it can be tuned via the relative orientation of $\mathbf{n}$ and $\mathbf{s}$, which also allows to separate these effects from other spurious contributions.

Let us first consider that $\dot{\mathbf{n}}=(T_\mathrm{mag}-T_\mathrm{N})=0$ and that the charge current bias applied to the NM induces a finite spin accumulation at the MOI/NM interface ($\mu_\mathrm{s}(0)\neq0$). The spin current over the interface is then governed by two contributions. In case of $\mathbf{n}\parallel\mathbf{s}$ only $g$ is relevant for $\mathbf{j}_\mathrm{s,int}$ resulting in a finite magnon accumulation underneath the NM. For $\mathbf{n}\perp\mathbf{s}$, $\tilde{g}^{\uparrow\downarrow}$ is the dominant contribution for $\mathbf{j}_\mathrm{s,int}$, which leads to a spin transfer torque acting on the magnetic order parameter. Due to the temperature dependence of $g$ one can assume that $|\tilde{g}^{\uparrow\downarrow}|\gg g$~\cite{Cornelissen2016}. Thus, the pure spin current across the interface is much larger for $\mathbf{n}\perp\mathbf{s}$ than for $\mathbf{n}\parallel\mathbf{s}$. The finite spin current flow across the interface causes a finite transverse pure spin current flow in the NM within the electron spin diffusion length. The finite transverse pure spin current flow leads to an effective increase in the path the electrons have to traverse for contributing to the charge current flow. From this, it follows that the transverse pure spin current flow effectively increases the resistance in the NM. Thus, the larger $\mathbf{j}_\mathrm{s,int}$ the larger the resistance increase. From this discussion we find that for $\mathbf{n}\parallel\mathbf{s}$ the resistance of the NM is smaller than for $\mathbf{n}\perp\mathbf{s}$. This resistance change is the spin Hall magnetoresistance (SMR) first observed in yttrium iron garnet/platinum heterostructures~\cite{Nakayama2013,althammer_quantitative_2013,chen_theory_2013}. The SMR has been employed to study various MOI/NM heterostructures~\cite{Nakayama2013,althammer_quantitative_2013,chen_theory_2013,Hahn2013SMR,Vlietstra2013,weiler_experimental_2013,Meyer2014,Han2014,Isasa2014,Aqeel2015,Wu2015,shang_pure_2016,hui_spin_2016,Avci2016,Vlez2016,hoogeboom_negative_2017,ji_spin_2017,hou_tunable_2017,manchon_spin_2017,Chen2016SMRReview,wang_comparative_2017}. In the last years, the SMR is prominently used for the investigation of antiferromagnetic insulators, which enables us to track the orientation of the N\'{e}el order parameter and to detect magnetic domains in this class of materials~\cite{Han2014,Manchon2017,hoogeboom_negative_2017,ji_spin_2017,hou_tunable_2017,manchon_spin_2017,Wang2017,Fischer2018,Ji2018,Baldrati2018,Schlitz2018,Baldrati2019,Lebrun2019,Geprgs2020,Das2021}. In addition, based on the SMR effect it is possible to quantify spin-orbit torque effects~\cite{Avci2016,Chen2018,Guo2019,Zhang2019,Cheng2020,Chen2020,Finley2020} and investigate topological spin textures in MOIs~\cite{aqeel_electrical_2016,Ahmed2019,Daniels2019,Ding2019,Cheng2019,Shao2019,Aqeel2021}.

For the all-electrical magnon transport experiments we consider the case that $\mathbf{s}$ and $\mathbf{n}$ are arranged collinear to each other. For this arrangement the interfacial spin-transfer torque acts on the thermally excited magnons in the MOI. Assuming an easy-axis antiferromagnetic order in the MOI, the two magnetic sublattices of the magnetic system give rise to two magnon modes (called here $\alpha$- and $\beta$-mode) with opposite chirality and thus carrying also opposite spin oriented along $\mathbf{n}$~\cite{Shen2019,Rezende2019,flebus2021magnonics}. This configuration is depicted in Fig.~\ref{figure:MagnonGeneration}. Due to the inelastic electron spin-flip scattering at the interface magnons of one spin orientation are generated (Fig.~\ref{figure:MagnonGeneration}(a)) and magnons of the opposite spin orientation are absorbed at the interface (Fig.~\ref{figure:MagnonGeneration}(b)). This leads to a non-equilibrium magnon accumulation for one magnon mode, while there is a magnon depletion of the other mode near the interface. This non-equilibrium accumulation and depletion of magnons drives diffusive magnon currents in the MOI, where we need to take into account the finite spin lifetime of the magnon system, and thus a finite magnon spin diffusion length.

Up to now, we only accounted for the charge-to-spin current conversion process, when driving a charge current through the NM. However, due to the finite resistance of the NM one also needs to account for Joule heating at the injector in these magnon transport experiments. (Fig.~~\ref{figure:MagnonGeneration}(c)) This locally increases the temperature of the MOI at the interface, which leads to additional heat currents flowing in the MOI. In this regard, heat will be conducted via excitations of the crystal and magnetic lattice, \ie~an additional phonon and magnon accumulation underneath the injector in the MOI originates from Joule heating. In contrast to the spin current injection via the charge-to-spin current conversion process in the NM, the magnon accumulation is independent of the orientation of $\mathbf{n}$ with respect to $\mathbf{s}$ and does not depend on the spin convertance $g$. For the antiferromagnetic MOI, this will result in an increase of magnons for both modes. In addition, we need to account for the fact that the interfacial spin Seebeck contribution $S$ will also lead to a spin current injection into the MOI, if there is a finite temperature difference between the electrons in the NM and the magnons in the MOI. This will again lead to an accumulation of magnons for one magnon mode and a depletion for the other, which now depends on the sign of $S$ and $T_\mathrm{mag}-T_\mathrm{N}$. As already demonstrated experimentally such a thermal spin-transfer torque leads to sizeable contributions in nanostructures based on NM/MOI heterostructures~\cite{safranski_spin_2017}.

\subsection{Magnon transport}
\begin{figure}
 \centering
 \includegraphics[width=85mm]{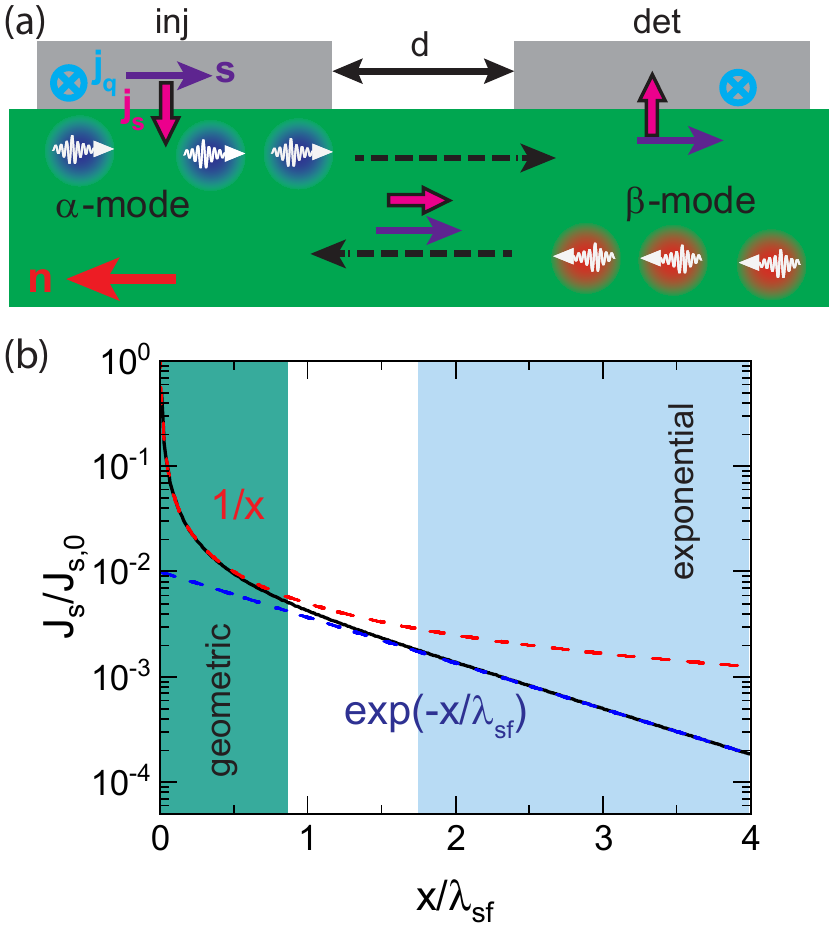}\\
 \caption[All-electrical magnon transport experiments]{(a) Schematic illustration of the all-electrical magnon transport experiments. A MOI (green area) with order parameter $\mathbf{n}$ is in contact with two strips of a NM (grey) referred to as injector (inj) and detector (det). When a charge current $\mathbf{j}_\mathrm{q}$ is applied to the injector, it is transformed into a spin current $\jsd$ flowing towards the HM/MOI interface with spin polarization $\mathbf{s}$. This induces an electron spin accumulation at the interface, and leads to an accumulation/depletion of thermal $\alpha$-mode/$\beta$-mode magnons, driving a diffusive magnon spin current in the MOI towards the detector strip. Utilizing the inverse process at the detector interface a spin current is injected into the NM and transformed into a charge current. This charge current can be electrically detected as an open circuit voltage. (b) Calculated evolution of the normalized magnon spin current $J_\mathrm{s}/J_\mathrm{s,0}$ as a function of the distance $x/\lspin$. The black line is the analytical expression obtained from Eq.(\ref{eq:spindiff}). The red dashed line corresponds to the $1/x$ approximation for short distances, while the blue dashed lines depicts the exponential decrease for large distances.}
  \label{figure:MagnonTransport}
\end{figure}

In the all-electrical magnon transport experiments we exploit the fact that in electrical conductors charge-to-spin current conversion processes allow to use a charge current $\mathbf{j}_\mathrm{q}$ to generate a spin current $\jsd$ flowing towards the NM/MOI interface. Within the geometry given in Fig.~\ref{figure:MagnonTransport} and using Eq.(\ref{eq:SHE}) the spin polarization $\mathbf{s}$ is fixed in direction, \ie~it is defined by the geometry itself and the direction of $\mathbf{j}_\mathrm{q}$. At the NM/MOI interface the flow of spin currents across the interface is determined via Eq.(\ref{equ:InterfacialSpinCurrent}).  Due to the interfacial coupling at the MOI/NM interface this then leads to an accumulation of one type of thermal magnons ($\alpha$-mode magnons in Fig.~\ref{figure:MagnonTransport}) and a depletion of the other type of thermal magnons ($\beta$-mode)underneath the injector strip. If one reverses the charge current direction, the spin polarization $\mathbf{s}$ is inverted and thus also the type of magnons that are accumulated/depleted underneath the injector are interchanged. In case of ferromagnetic/ferrimagnetic order in MOIs only one magnon mode is available (for example just the $\alpha$-mode magnons prevail). Thus, we can describe the process as an accumulation/depletion of thermal magnons for just one magnon mode. As discussed in more detail in the following, we then obtain a diffusive spin current carried by magnons in the MOI, transporting spin current with polarization $\mathbf{s} \parallel -\mathbf{n}$ towards the detector strip. At the detector, the magnon spin current pumps a spin current across the interface into the NM. In the NM detector this spin current $\jsd$ is transformed into a charge current $\mathbf{j}_\mathrm{q}$ via spin-to-charge current conversion processes. Under electrical open circuit conditions, this charge current then leads to a charge accumulation at the end of the NM strip and can be electrically detected as a voltage drop across the NM strip. It is important to note that within this spin current transport process the NM detector strip acts as a spin current sink, \ie~the magnon spin current is absorbed by the NM strip. This is achieved by the finite spin diffusion length in the NM and MOI, a direct cause of the limited lengthscale upon which spin angular momentum is conserved. Thus, the thickness of the NM strip should exceed the value of $\lspinNM$ of the NM to maximize the amount of spin current flow. Moreover, since the experiments require thermal fluctuations of the magnetic lattice to achieve finite spin current transport, these experiments require finite temperature and a large value of $g$ (the spin conductance governing the magnitude of the interfacial spin current). Another important aspect is the fact that magnon transport is only possible if $\mathbf{n}$ is collinear with $\mathbf{s}$. This allows to disentangle magnon transport signals from any additional contributions in the experiment by varying the orientation of $\mathbf{n}$ via an external control parameter, such as the orientation of the applied external magnetic field.

In addition we need to account for thermal effects associated with Joule heating due to the finite resistance of the injector strip. We here assume that contribution from the heat current transport across the interface dominate over the interfacial spin Seebeck effect. The generated magnon accumulation also diffuses towards the detector strip. In case of an antiferromagnetic MOI, there are two magnon modes with opposite spin polarization and one expects that these two contributions compensate each other, if we assume equal properties of these two degenerate magnon modes. This compensation effect can be lifted by introducing a deviation from the collinear arrangement of the magnetic moments of the two magnetic sublattices, \ie~by introducing a finite net magnetization for example by the application of an external magnetic field~\cite{Klaui2018,Rezende2019}. For a ferromagnetic MOI, we only have a single magnon mode, thus at the detector $\mathbf{s} \parallel -\mathbf{n}$ for the magnons injected via Joule heating for the electrical detection. This correlation of $\mathbf{s}$ with $\mathbf{n}$ in a ferromagnetic material in combination with the inverse spin Hall effect detection in the NM allows to disentangle this thermal contributions from other spurious effects by varying the orientation of $\mathbf{n}$ in the experiment. We note that in addition temperature differences between the electrons and magnons at the detector interface can also cause an addition spin current injection into the NM detector strip. Thus. a full quantitative understanding of all signal contributions for the thermal part is very challenging and different results have been observed in the experiment~\cite{shan_influence_2016,GomezPerez2020_magnon,an2020shortrange}

Up to now, we only accounted for the electrical injection and detection mechanisms in the NM to explain the all-electrical magnon transport experiments. Since the injected pure spin current is transported via magnons in the MOI, we also need to account for the magnon transport properties, the following discussion is based on the more detailed description of magnon transport in Refs.~\cite{Cornelissen2016,Shen2019}. In the following we assume the case for two sublattice antiferromagnetic insulators (AFIs) with the two magnon modes $\alpha$ and $\beta$ carrying opposite spin momentum of $\pm \hbar$~\cite{Rezende2016,Rezende2019,Shen2019,Troncoso2020}. In this way, we can consider spin current transport in AFIs as the magnonic analogue of electron spin transport, where also two opposing spin directions contribute. In case of AFIs, the spin direction of the two magnon modes is collinear to the orientation of the N\'{e}el vector. These two modes are energetically degenerate with a dispersion relation for each mode following $E^{\alpha(\beta)}_\mathrm{AFI}(\mathbf{k})=\sqrt{E_0^2+J_\mathrm{ex}^2\mathbf{k}^2}$, with the $E_0$ the magnon gap energy and $J_\mathrm{ex}$ the exchange stiffness parameter. Within the Boltzmann equation of motion, we describe the magnons in each mode via a distribution function $f^{\alpha(\beta)}(\mathbf{r}, \mathbf{k}, t)$ depending on spatial coordinate $\mathbf{r}$, momentum $\mathbf{k}$ and time $t$:
\begin{equation}
\frac{\partial f^{\alpha(\beta)}}{\partial t}+\frac{1}{\hbar}\frac{\partial E^{\alpha(\beta)}_\mathrm{AFI}(\mathbf{k})}{\partial \mathbf{k}}\nabla f^{\alpha(\beta)}=\Gamma^{\alpha(\beta)}_\mathrm{scat}+\Gamma^{\alpha(\beta)}_\mathrm{int}\;,
\label{eq:MagnonBoltzmannAFI}
\end{equation}
with $\Gamma^{\alpha(\beta)}_\mathrm{scat}$ the relaxation rate due to magnon scattering and $\Gamma^{\alpha(\beta)}_\mathrm{int}$ the relaxation (or injection) rate due to interfacial effects, i.e. describing the contributions from the NM/MOI interface defined in Eq.(\ref{equ:InterfacialSpinCurrent}). For the following discussion we treat $\Gamma^{\alpha(\beta)}_\mathrm{int}$ as a small perturbation of the system and focus on $\Gamma^{\alpha(\beta)}_\mathrm{scat}$. For the magnon system we need to account for multiple scattering processes and assume a scattering relaxation time approach:
\begin{equation}
\Gamma^{\alpha(\beta)}_\mathrm{scat}=-\sum_i \frac{f^{\alpha(\beta)}-\overline{f}^{\alpha(\beta)}_i}{\tau_i}\;.
\label{eq:ScatteringRelax}
\end{equation}
Here, $\tau_i$ is the relaxation time for scattering process $i$ and $\overline{f}^{\alpha(\beta)}_i$ is the quasi-equilibrium distribution function the magnon system relaxes into. In general, $\tau_i$ depends on momentum, for simplicity we here assume a constant value. For the different scattering processes we follow the discussion in Ref.~\cite{Shen2019} and distinguish between magnon number conserving and magnon number non-conserving processes. The former can be split into magnon-magnon scattering, described via $\tau_\mathrm{mm}$ and enabling energy transfer within and between the two magnon modes, magnon-phonon scattering, described by $\tau_\mathrm{mp}$ allowing for energy transfer to the phonon bath, and elastic magnon scattering with defects $\tau_\mathrm{el}$. For the magnon number non-conserving processes we need to differentiate in the AFI between two different contributions. Magnon-magnon scattering processes, which maintain the difference in magnon numbers between the two magnon modes parameterized by $\tau_\mathrm{mm,nc}$. In other words, pairs of magnons, one magnon for each mode, can be generated or annihilated in a single magnon scattering event~\cite{Shen2019}. The remaining processes that do not conserve the number of magnons are described by the magnon relaxation rate $\tau_\mathrm{mr}$. The finite $\tau_\mathrm{mr}$ also reflects the fact that the magnon number is not a conserved transport quantity, similar as the spin in electron transport. As a next step we assume that $\tau_\mathrm{mr}\gg\tau_\mathrm{mp},\tau_\mathrm{mm},\tau_\mathrm{mm,nc},\tau_\mathrm{el}$, which is a reasonable approximation for all investigated MOIs so far~\cite{Cornelissen2016,Shen2019}. In good approximation the magnon number is thus conserved on a characteristic lengthscale $\lspin$ in the AFI. To solve Eq.(\ref{eq:MagnonBoltzmannAFI}) we now assume a linearized solution for the distribution function 
\begin{equation}
    f^{\alpha(\beta)}=g^{\alpha(\beta)}(\mathbf{k})+n^{\alpha(\beta)}_\mathrm{B}(E^{\alpha(\beta)}_\mathrm{AFI}(\mathbf{k}),\mu^{\alpha(\beta)}_\mathrm{mag},T^{\alpha(\beta)})\,,
    \label{eq:LinDistribution}
\end{equation} 
where $n^{\alpha(\beta)}_\mathrm{B}=(\exp[(E^{\alpha(\beta)}_\mathrm{FMI}(\mathbf{k})-\mu^{\alpha(\beta)}_\mathrm{mag})(k_\mathrm{b}T^{\alpha(\beta)})^{-1}]-1)^{-1}$ is the Bose-Einstein distribution function with a locally defined magnon chemical potential $\mu^{\alpha(\beta)}_\mathrm{mag}(\mathbf{r})$ and temperature $T^{\alpha(\beta)}(\mathbf{r})$ for each magnon mode in the AFI. The magnon number conserving magnon-magnon scattering processes lead to an efficient energy exchange between the two modes such that temperature equilibrium  $T^{\alpha}=T^{\beta}$ is reached on the timescale of $\tau_\mathrm{mm}$. Moreover, the magnon-magnon scattering processes summarized via $\tau_\mathrm{mm,nc}$ lead to $\mu^{\alpha}_\mathrm{mag}=-\mu^{\beta}_\mathrm{mag}$. This remarkable condition is similar to the case of electron spin transport in the two spin channel model. However, for electron transport the connection between spin chemical potential for up- and down-spin electrons arises from local charge neutrality (maintaining the charge free nature of a pure spin current), while for magnons in AFIs it is a direct consequence of magnon-magnon scattering. This fact has implications for the electrical generation/absorption at the NM/AFI interface, since the interface process now only needs to couple efficiently to one of the two magnon modes and magnon-magnon scattering ensures that both magnon modes will be affected by the pure spin current injection.

Substituting Eq.(\ref{eq:LinDistribution}) into Eq.(\ref{eq:MagnonBoltzmannAFI}) we get an expression for $g(\mathbf{k})$
\begin{align}
g^{\alpha(\beta)}(\mathbf{k})&=\tau\left(-\frac{\partial n^{\alpha(\beta)}_\mathrm{B}(E^{\alpha(\beta)}_\mathrm{AFI}(\mathbf{k}),\mu^{\alpha(\beta)}_\mathrm{mag},T^{\alpha(\beta)})}{\partial E^{\alpha(\beta)}_\mathrm{AFI}(\mathbf{k})}\right) \nonumber\\ & \frac{\partial E^{\alpha(\beta)}_\mathrm{AFI}(\mathbf{k})}{\partial \mathbf{k}} 
\left[ -\nabla \mu^{\alpha(\beta)}_\mathrm{mag}-\frac{E^{\alpha(\beta)}_\mathrm{AFI}(\mathbf{k})-\mu^{\alpha(\beta)}_\mathrm{mag}}{T^{\alpha(\beta)}}\nabla T^{\alpha(\beta)}\right]\;,
\label{eq:AFI_gDistribution}
\end{align}
with $(\tau)^{-1}=\sum (\tau_i)^{-1}$ and neglecting higher order contributions in $g^{\alpha(\beta)}$. With Eq.(\ref{eq:AFI_gDistribution}), we now can define the magnon particle current density for each magnon mode
\begin{equation}
\mathbf {j}^{\alpha(\beta)}_{\mathrm{mag}} =  \int \frac{d\mathbf{k}}{(2\pi)^3} g^{\alpha(\beta)}(\mathbf{k}) \frac{\partial E^{\alpha(\beta)}_\mathrm{AFI}(\mathbf{k})}{\partial \mathbf{k}}\;.
\label{eq:AFI_MagnonParticleCurrent}
\end{equation}
For each magnon mode we also find a spin current density (each magnon carries a spin moment of $\pm \hbar$) 
\begin{equation}
\mathbf {j}^{\alpha(\beta)}_{\mathrm{s}} = \pm \hbar \int \frac{d\mathbf{k}}{(2\pi)^3} g^{\alpha(\beta)}(\mathbf{k}) \frac{\partial E^{\alpha(\beta)}_\mathrm{AFI}(\mathbf{k})}{\partial \mathbf{k}}\;.
\label{eq:AFI_MagnonSpinCurrent}
\end{equation}
In similar fashion, we can define the magnon heat current density in the AFI for each mode (each magnon carries energy $[E_\mathrm{AFI}(\mathbf{k})-\mu_\mathrm{mag}]$)
\begin{equation}
\mathbf {j}^{\alpha(\beta)}_{\mathrm{h}} = \int \frac{d\mathbf{k}}{(2\pi)^3} g^{\alpha(\beta)}(\mathbf{k}) (E^{\alpha(\beta)}_\mathrm{AFI}(\mathbf{k})-\mu^{\alpha(\beta)}_\mathrm{mag})\frac{\partial E^{\alpha(\beta)}_\mathrm{AFI}(\mathbf{k})}{\partial \mathbf{k}}\;.
\label{eq:AFI_MagnonHeatCurrent}
\end{equation}
These current are driven by the gradients in $\mu^{\alpha(\beta)}_\mathrm{mag}$ and $T^{\alpha(\beta)}$. In similar fashion as for electron spin transport, we define the total magnon particle current $\mathbf {j}_{\mathrm{mag}}=\mathbf {j}^{\alpha}_{\mathrm{mag}}+\mathbf {j}^{\beta}_{\mathrm{mag}}$, the total magnon spin current $\mathbf {j}_{\mathrm{s}}=\mathbf {j}^{\alpha}_{\mathrm{s}}+\mathbf {j}^{\beta}_{\mathrm{s}}$, and the total magnon heat current $\mathbf {j}_{\mathrm{h}}=\mathbf {j}^{\alpha}_{\mathrm{h}}+\mathbf {j}^{\beta}_{\mathrm{h}}$. Moreover, we define the total magnon chemical potential $\mu_{\mathrm{mag}}=\mu^{\alpha}_{\mathrm{mag}}+\mu^{\beta}_{\mathrm{mag}}$, the magnon spin chemical potential $\mu_{\mathrm{mag,s}}=\mu^{\alpha}_{\mathrm{mag}}-\mu^{\beta}_{\mathrm{mag}}$ and assume fast thermalization between the two magnon modes such that we can define the magnon temperature $T=T^\alpha=T^\beta$. Then we can write down the generalized transport equation of the magnon system
\begin{equation}
\left (
\begin{array}{c}\mathbf j_\mathrm{mag} \\\mathbf j_\mathrm{h} \\ \mathbf {j}_{\mathrm{mag,s}}
\end{array} \right ) =
\left( \begin{array}{ccc}
L^{11} & L^{12} & L^{13}\\
L^{21} & L^{22} & L^{23}\\
L^{31} & L^{32} & L^{33}
\end{array} \right) \left (
\begin{array}{c}\nabla \mu_{\mathrm{mag}} \\ -\nabla T /T \\ \nabla \mu_{\mathrm{mag,s}}
\end{array} \right )\;.
\label{eq:transport_magnon}
\end{equation}
Here, $L^{ij}$ are the linear response matrix elements determined via integration in $\mathbf{k}$-space. Moreover, it holds that $L^{ij}=L^{ji}$ for $i\neq j$ due to Onsager reciprocity. We see that a spatial gradient in $\mu_{\mathrm{mag,s}}$ as well as in $T$ will drive a spin current as well as a heat current. For a full description of the transport we now need to define the continuity equations for \jsd and \jhd, which also accounts for the intrinsic loss rate of magnons in the MOI and thus introduces the characteristic transport lengths of the system. This is in contrast to charge current transport, were we do not have to account for any losses in the system, such that charge is a conserved transport quantity. We can write down the continuity equations in the following simplified form~\cite{Cornelissen2016}
\begin{align}
\left (
\begin{array}{c} 
\frac{\partial \rho_\mathrm{mag}}{\partial t} + \nabla \mathbf j_\mathrm{mag} \\
\frac{\partial Q_\mathrm{mag}}{\partial t}+ \nabla \jhd \\ 
\frac{\partial \rho_\mathrm{mag,s}}{\partial t} + \nabla \jsd
\end{array} \right ) &= -
\left( \begin{array}{ccc}
\Gamma_{\rho \mu}& \Gamma_{\rho T} & \Gamma_{\rho \mu_\mathrm{s}} \\
\Gamma_{Q \mu} & \Gamma_{Q T} & \Gamma_{Q \mu_\mathrm{s}} \\
\Gamma_{\rho_\mathrm{s} \mu} & \Gamma_{\rho_\mathrm{s} T} & \Gamma_{\rho_\mathrm{s} \mu_\mathrm{s}}
\end{array} \right) \nonumber \\ & \left (
\begin{array}{c} \mu_{\mathrm{mag}} \frac{\partial \rho_\mathrm{mag}}{\partial \mu_{\mathrm{mag}}} \\ T \frac{\partial Q_\mathrm{mag}}{\partial T} \\ \mu_{\mathrm{mag,s}} \frac{\partial \rho_\mathrm{mag,s}}{\partial \mu_{\mathrm{mag,s}}}
\end{array} \right )\;,
\label{eq:continuityequation_magnon}
\end{align}
with $\rho_\mathrm{mag}$, $Q_\mathrm{mag}$, and $\rho_\mathrm{mag,s}$ the non-equilibrium magnon density, magnon heat density, and magnon spin density, respectively. $\Gamma_{ij}$ are the (local) relaxation and generation rates caused by magnon scattering and spin-/heat injection into the AFI, \ie consisting of $\Gamma^{\alpha(\beta)}_\mathrm{scat}$ and $\Gamma^{\alpha(\beta)}_\mathrm{int}$. Due to the cross terms in Eqs.(\ref{eq:transport_magnon}),(\ref{eq:continuityequation_magnon}) $\mu_\mathrm{mag}$, $T$, and $\mu_\mathrm{mag,s}$ contribute to $\jsd$ and $\jhd$. On short timescales (few ps), magnon-phonon scattering and magnon-magnon scattering lead to the fact that the non-equilibrium magnon temperature profile quickly relaxes to the phonon temperature profile and $\mu_\mathrm{mag}$ approaches 0 ($\mu^{\alpha}_{\mathrm{mag}}=-\mu^{\beta}_{\mathrm{mag}}$ due to energy exchange between the two magnon modes) on few nm lengthscales~\cite{Cornelissen2016,Shen2019}. In contrast, magnon spin non-conserving scattering happens at much longer timescales of the order of several $100\,\mathrm{ns}$. Taking this into account, we can simplify our model of magnon spin transport and neglect $\nabla \mu_\mathrm{mag}$ and $\nabla T$ as a driving force for $\jsd$ and only account for $\nabla \mu_{\mathrm{mag,s}}$, thus we obtain the magnon spin diffusion equation by combining Eqs.(\ref{eq:transport_magnon}),(\ref{eq:continuityequation_magnon})
\begin{equation}
D_\mathrm{mag,s} \mathbf{\nabla}^2 \mu_{\mathrm{mag,s}} = \frac{\mu_{\mathrm{mag,s}}}{\tau_\mathrm{s}}\;.
\label{eq:spindiff}
\end{equation}
Here, $D_\mathrm{mag,s}$ is the magnon spin diffusion constant and $\tau_\mathrm{s}$ the magon spin lifetime, \ie~the magnon spin non-conserving scattering times. We then define the magnon spin diffusion length $\lspin=\sqrt{D_\mathrm{mag,s}\tau_\mathrm{s}}$. In Fig.~\ref{figure:MagnonTransport}(b), we illustrate the normalized magnon spin current as a function of the spatial coordinate $x$ for the case of a strictly 1d diffusion process (the only scenario, were analytical expressions are possible). For this graph, we assumed that a fixed spin current is injected at $x=0$ and the magnon spin current vanishes at $x=\inf$ in the AFI. The black curve is the obtained analytical expression ($\jsd\propto \exp(x/\lspin)/(1-\exp(2x/\lspin))$), while the dashed red and blue line are approximations for certain distance regimes. For $x\ll\lspin$ (from Fig.~\ref{figure:MagnonTransport}(b) this is the regime $x\leq0.5\lspin$), we see that $\jsd$ reduces as $1/x$, which is identical to a diffusion process without relaxation and thus equivalent to diffusive electron charge transport. Within the short distance regime, we can then describe the magnon spin transport in very good approximation like a charge current transport and associate a magnon spin conductance $\sigma_\mathrm{mag,s}$ to describe the linear relation between $\jsd$ and $\nabla \mu_\mathrm{mag,s}$ under the assumption of a localized injection~\cite{Cornelissen2016}. For $x\gg\lspin$ (corresponding to $x\geq 1.5\lspin$), we obtain an exponential dependence of $\jsd$, which enables a distance dependent evaluation of $\lspin$ in the experiment.

Lets briefly review this model of all-electrical magnon transport in AFIs: In case of the charge-to-spin current conversion driven spin current injection from the NM into the AFI, the inelastic spin-flip scattering process of electrons at the NM/AFI interface enables to generate/absorb magnons over the thermally broadened energy scale $k_\mathrm{b}T$ in the FMI~\cite{Bender2015}. Joule heating on the other hand dominantly generates magnons with energy close to $k_\mathrm{b}T$. Energy relaxation of these injected magnons happens on short timescales within less than a few ps, and the quasi-equilibrium is achieved by very efficient magnon-magnon scattering events maintaining the spin polarization in the system. On these timescales, there is a strong interaction between the magnons, while still maintaining the injected spin polarization. Further magnon scattering processes then lead to a decay of spin polarization on the timescale of $\approx 10-100\;\mathrm{ns}$. At present it is not fully understood what the dominant scattering contribution is for this relaxation into thermal equilibrium and most certainly the details will depend on the material. Possible scenarios include magnon-phonon scattering or magnon scattering with (magnetic) defects (magnetic domain structure). For the prototype AFI $\alpha$-Fe$_2$O$_3$ (hematite) we find  $D_\mathrm{mag}\approx3\times10^{-5}\;\mathrm{m^2s^{-1}}$ and $\tau_\mathrm{s}\approx10\;\mathrm{ns}$ from all-electrical magnon transport experiments in thin film hematite within the easy-plane phase~\cite{Han2020}.

One of the most remarkable properties of magnon transport in AFIs are the two magnon modes with opposite spin, resembling the magnonic analog of electron spin transport. In principle, a coherent coupling between the two magnon modes is possible via magnetic anisotropy, Dzyaloshinskii-Moriya interaction, or dipolar magnetic fields. In this way magnon transport in AFIs could possibly exhibit magnonic equivalents of electronic spin transport effects. There is a growing number of theoretical concepts already out exploiting this equivalent nature, predicting for example the existence of topological magnon states in AFIs or emergent spin-orbit coupling effects in magnon transport~\cite{Cheng2016,Cheng2016B,Mook2017,Daniels2018,Hou2019,Kawano2019,Kawano2019B,Daniels2019,Shen2020,kamra_antiferromagnetic_2020}. First experiments on magnon transport via all-electrical means in AFIs were fully compatible with a simple diffusive magnon spin transport, where the injected spin orientation is maintained within the AFI~\cite{Klaui2018,Klaui2020,Han2020,Lebrun2020}. However, within the last year new experiments emerged, that confirm the electronic transport analogy via the observation of the magnon Hanle effect in easy-plane antiferromagnetic insulators~\cite{wimmer_observation_2020,kamra_antiferromagnetic_2020,Ross2020}. These advances will be discussed in section \ref{sec:magnonPseudo} and enable the manipulation of the transported spin direction for AFIs.
Thus, magnon transport in AFIs has the potential to become a resourceful playground to realize the magnonic equivalents of electron transport phenomena. In this regard, all-electrical magnon transport experiments provide a unique and simple access as it requires simple dc-charge current transport experiments as compared to the more complex excitation of magnons in AFIs via high frequency electro-magnetic radiation (starting in the range of several $100\;\mathrm{GHz}$). Moreover, in case of the SHE-based spin current injection from the NM into the AFI, we can control the direction of the injected spin polarization via the direction of charge current flow and thus controlling generation/annihilation of $\alpha$-, $\beta$-magnons in the AFI.

For ferromagnetic insulators, the situation is quite similar to the case of AFIs. Now we need to account for one magnon mode with a parabolic dispersion relation carrying a spin of $\hbar$. One then finds the magnon chemical potential $\mu_\mathrm{mag}$, which is equal to $\mu_\mathrm{mag,s}$ for ferromagnetic insulators. The dominant magnon scattering processes in the FMI then lead to a fast energy relaxation of the injected magnons, while the spin polarization is maintained on much longer timescales. In the prototype FMI yttrium iron garnet (YIG) at room temperature, magnon-magnon scattering (dominated by 4-magnon scattering processes) and magnon-phonon scattering are important for this energy relaxation process. Such that injected magnons will relax towards the minimum in energy of the magnon dispersion relation. In more detail, in Ref.~\cite{Cornelissen2016} the relaxation times for thin film YIG have been calculated and they obtain at room temperature: $\tau_\mathrm{mr}\approx10-100\,\mathrm{ns}$, $\tau_\mathrm{mp}\approx0.1-1\,\mathrm{ps}$, $\tau_\mathrm{mm}\approx0.1-1\,\mathrm{ps}$, and $\tau_\mathrm{el}\approx10-10^5\,\mathrm{ps}$. From these calculations it is evident that the magnon-conserving processes dominate magnon scattering by at least 2-3 orders of magnitude. In this case, scattering on defects in YIG is not the important factor for the lengthscale of ballistic transport (for YIG $<1\,\mathrm{nm}$ at room temperature~\cite{Bender2015}), but we are rather limited by the strong interaction of magnons with phonons and themselves. Neglecting a $\mathbf{k}$ dependence one can estimate $\tau_\mathrm{s}\approx(\alpha_\mathrm{G} k_\mathrm{b} T/\hbar)^{-1}$ from the Gilbert damping constant $\alpha_\mathrm{G}$ of the ferromagnetic resonance in the FMI. For thin film YIG one extracts $D_\mathrm{mag}\approx2\times 10^{-4}\;\mathrm{m^2s^{-1}}$ and $\tau_\mathrm{s}\approx500\;\mathrm{ns}$ at room temperature from the magnon transport experiments~\cite{Cornelissen2016}. Another important aspect for spin transport via magnons in FMIs is that the spin polarization direction of the magnon spin current is determined by the orientation of the magnetization in the FMI. Thus, this spin direction is maintained within a FMI with homogenous magnetization orientation, such that the injected spin direction is maintained, when transported through the FMI via magnons.

In the all-electrical magnon transport experiments with an injector and detector NM strip, we typically apply a few volts to the injector to drive a charge current through the injector strip. The magnon transport voltage signal at the detector strip is in the range of a few $100\;\mathrm{nV}$). From an application point of view, we can thus treat all-electrical magnon transport experiments as a voltage conversion process with a specific efficiency. One thus finds a reduction from the applied voltage to the detected voltage by 6-7 orders of magnitude. This provides a major challenge to overcome for practical applications and it is important to understand the origin of the voltage conversion efficiency. Taking aside the influence of the length of the Pt strips (geometric scaling), the efficiency of the voltage conversion process is governed by three main contributions~\cite{Cornelissen2016}. First, the efficiency of the charge-to-spin conversion process in the NM, which enters quadratically as this process takes place twice: conversion of charge current to spin current in the injector and conversion of the spin current to a charge current in the detector. For Pt as the NM, $\thetaSH\approx 0.1$, such that this leads to reduction by two orders of magnitude. Next, the transparency of the NM/MOI interface governed by $g$ is important for the injection of the spin current into the MOI. This contribution enters again quadratically since the spin current transport goes through the interface at the injector and at the detector strip. Assuming that $g\approx0.1\tilde{g}_r^{\uparrow\downarrow}$ (reasonable assumtion for yttrium iron garnet at room temperature~\cite{Cornelissen2016}), $\tilde{g}_r^{\uparrow\downarrow}=1\times10^{19}\;\mathrm{m^{-2}}$~\cite{weiler_experimental_2013} and typical values for the spin diffusion length and resistivity of Pt, one finds a transparency of about $10^{-2}$, and thus this contribution leads to a total reduction by 4 orders of magnitude. Finally, the magnon spin transport in MOI is also relevant for the magnitude of the detector voltage. This contribution can be significantly reduced by using a distance between injector and detector strip much shorter than $\lspin$. As evident from this discussion the most dominant contribution for the efficiency of the voltage conversion is mostly determined by the transparency of the NM/MOI interface for spin currents. Generally, this requires an increase in $g$ and $\tilde{g}_r^{\uparrow\downarrow}$, an increase in resistivity of the NM and the electron spin diffusion length $\lspinNM$ in the NM. Moreover, further enhancement can be achieved by increasing the spin-to-charge conversion efficiency $\thetaSH$ in the NM.  The spin conductance $g$ is maximized at temperatures close to the magnetic ordering temperature of the MOI. A first temperature dependent study in yttrium iron garnet/platinum structures reaching the ordering temperature~\cite{schlitz2020nonlocal} only yielded an improvement by a factor of two. However, in these experiments it also became apparent that due to the elevated temperatures interdiffusion processes at the interface significantly reduced the spin transparency of the interface~\cite{Bai2019_YIG,Liu2020_spinpump}. Interestingly, recent non-local spin transport experiments in 2d quantum magnets show that the voltage at the detector is only reduced by 3 orders of magnitude~\cite{Stepanov2018,Fu2021}, most likely caused by the improved spin transparency as the NM/MOI interfaces are defined within the 2d material itself.

It is also important to note that similar structures as discussed here have already been investigated by Kajiwara \textit{et al.}~\cite{Kajiwara:2010ff} in 2010. The main difference between these experiments and the all-electrical magnon transport experiments discussed in this article is that Kajiwara \textit{et al.} utilized Pt injector and detector electrodes separated by $1\;\mathrm{mm}$, a distance much larger than the typical magnon spin diffusion length in YIG ($\approx 10 \mathrm{\mu m}$~\cite{Cornelissen2015}). In this way, they only observed a non-linear detector voltage response to the current applied to the injector as they operated in the regime, where auto-oscillations are driven via the charge current in the YIG layer. In contrast, we find for the all-electrical magnon transport a direct linear response of the detector voltage to the applied charge current in the injector~\cite{Cornelissen2015,Goennenwein2015}. The signal Kajiwara \textit{et al.} observed is most likely influenced by non-diffusive spin transport via magnons. It is important to note that this non-linear response regime has been also studied in all-electrical magnon transport experiments~\cite{Thiery_2018}. We cover the aspect of magnon damping compensation and the associated auto-oscillations in MOIS via a SHE based spin current injection from an adjacent NM for all-electrical magnon transport in the following section.

\section{Electrical manipulation of magnon transport}

\begin{figure}
 \centering
 \includegraphics[width=85mm]{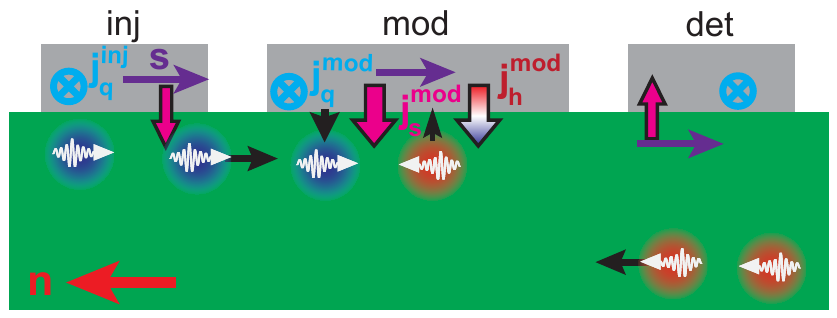}\\
 \caption[All-electrical magnontransport manipulation]{Illustration of all-electrical magnon transport experiments with charge current control of magnon transport. On top of the MOI three NM strips alow for the all-electrical magnon transport experiments. A charge current $\mathbf{J}_\mathrm{q}^{\mathrm{inj}}$ applied to the injector strip (inj) injects a pure spin current $\mathbf{J}_\mathrm{s}^{\mathrm{inj}}$ into the MOI. In the MOI this pure spin current is diffusively transported to the detector strip (det), where the arriving spin current $\mathbf{J}_\mathrm{s}^{\mathrm{det}}$ is transformed into a charge current $\mathbf{J}_\mathrm{q}^{\mathrm{det}}$ to electrically detect the magnon spin current. A charge current $\mathbf{J}_\mathrm{q}^{\mathrm{mod}}$ applied to the modulator NM strip (mod) injects an additional spin current $\mathbf{J}_\mathrm{s}^{\mathrm{mod}}$ into the MOI. This local spin current injection leads to a change in the magnon relaxation rate underneath the modulator in turn also affecting the magnon transport from injector to detector.}
  \label{figure:MagnonTransistor}
\end{figure}

In the following we discuss how the electrical magnon generation/absorption via spin current injection from an adjacent NM can be exploited to control the magnon transport within the MOI. This requires in the lateral magnon transport geometry with NM injector and detector an additional third NM strip, which we will refer to as the modulator strip (see Fig.~\ref{figure:MagnonTransistor}). By passing a charge current through the modulator we locally inject a pure spin current via charge-to-spin current conversion processes in the NM into the MOI. In addition, we need to account for Joule heating effects in the modulator causing local heating in our structure. We first discuss the limit of low charge currents applied to the modulator and then discuss effects associated with larger modulator charge currents.

\subsection{Charge current control of magnon transport}

As we discussed in the previous section, a charge current applied to the modulator allows to inject/absorb magnons via the SHE in the NM and inelastic electron spin-flip scattering, but also Joule heating in the modulator leads to an injection of magnons. These contributions can be treated in our transport model as a locally confined $\Gamma^{\alpha(\beta)}_\mathrm{int}$ interface injection/absorption rate, which depends on the charge current applied to the modulator. This change in $\Gamma^{\alpha(\beta)}_\mathrm{int}$ in combination with the efficient energy relaxation of the magnon system thus influences the magnon transport between injector and detector, because the magnon loss rate in the system will be changed via the additional magnon injection/absorption at the modulator interface. First pioneering experiments in this regard were conducted in Ref.~\cite{Cornelissen2018} with YIG/Pt heterostructures, utilizing $210\;\mathrm{nm}$ thin YIG films. In these experiments they observed small changes, but predicted sizeable effects for thinner YIG layers.

The magnon injection/absorption at the modulator is controlled via the applied charge current. We first focus on charge-to-spin current conversion processes in the NM, which allow to couple to the thermal fluctuations of the magnetic lattice of the MOI. Thus, magnons are injected or absorbed at the modulator interface. To simplify the discussion, we here assume that the SHE is the only contribution to the charge-to-spin current conversion processes in the NM. In this way, we find that for FMIs the direction and magnitude of charge current flow in the modulator and the orientation of $\mathbf{m}$ to the modulator define the sign and strength of $\Gamma_\mathrm{int}$. In case of AFIs, we need to account for the two different magnon modes and the fact that the spin accumulation will couple with opposite sign and possibly different magnitude to these two modes. Such that we have to define different interface absorption/relaxation rates for the two magnon modes $\Gamma^{\alpha(\beta)}_\mathrm{int}$. However, for both cases (AFIs and FMIs) we expect that $\Gamma_\mathrm{int}$ varies linearly with the applied charge current to the modulator $\mathbf{J}_\mathrm{q}^{\mathrm{mod}}$ due to the SHE. In addition, magnon diffusion processes will also lead to changes in the local magnon density at the injector influencing the interfacial spin current injection at the injector. Taking both these effects into account, we expect for the magnon transport signal (related to the voltage at the detector) obtained between the injector and detector a linear dependence on $\mathbf{J}_\mathrm{q}^{\mathrm{mod}}$ via SHE controlled magnon injection/relaxation.

In case of the additional Joule heating at the modulator due to the applied $\mathbf{J}_\mathrm{q}^{\mathrm{mod}}$ to the NM strip, the results on the magnon transport are more complex. On the one hand the additional heating will lead to a temperature increase in the MOI, which by itself will change magnon transport parameters in the MOI (like for example $D_\mathrm{mag}$, $\Gamma_\mathrm{scat}$) and affect the spin convertance $g$ at the injector. In addition, thermal gradients will lead to a local injection of magnons at the modulator affecting $\Gamma_\mathrm{int}$. Taking these effects into account, we expect a quadratic dependence on $\mathbf{J}_\mathrm{q}^{\mathrm{mod}}$ via Joule heating in the modulator for the magnon transport between injector and detector and thus the corresponding voltage signal at the detector. One problem also arising in these magnon transport experiments is that the electrical conductivity of the MOI will also depend on temperature and can lead to a leakage charge current flowing from injector to detector, which needs to be accounted for in the experiments~\cite{Thiery2018}. In addition, these charge current will generate local Oersted fields, which can also affect magnon transport properties~\cite{Cramer_2019_mt}.

As evident from this discussion, we can treat the induced changes on magnon transport between injector and detector for small $\mathbf{J}_\mathrm{q}^{\mathrm{mod}}$ applied to the modulator as small perturbations and expect changes in the magnon transport signal, which depend linearly and quadratically on $\mathbf{J}_\mathrm{q}^{\mathrm{mod}}$. Interestingly, at the limit of low MOI thicknesses and large values of $\mathbf{J}_\mathrm{q}^{\mathrm{mod}}$ it is possible to compensate the intrinsic magnon spin relaxation rate $(\tau_\mathrm{s})^{-1}$ locally at the modulator, leading to more drastic changes in the magnon transport. We discuss experiments in this regime in more detail in the following.

\subsection{Measurement techniques}
We want to briefly discuss the measurement techniques that can be used to experimentally investigate all-electrical magnon transport in the three strip magnon modulator geometry (see Fig.~\ref{figure:MagnonTransistor}). As already discussed, the experiments rely on driving a charge current through the injector to achieve magnon transport to the detector. At the detector the magnon transport can be detected as a voltage drop $V_\mathrm{det}$ along the NM strip. In the magnon transistor configuration we face the challenge that a charge current is driven through the injector ($I_\mathrm{inj}$) as well as the modulator($I_\mathrm{mod}$), thus $V_\mathrm{det}$ consists of contributions from injector and modulator. To separate the magnon transport signals from the injector and modulator one has to employ a charge current modulation technique. We here discuss two commonly used modulation techniques. 

One possibility is to drive a dc charge current through both injector and modulator. In order to separate in this case the different contributions to $V_\mathrm{det}$ from injector and detector, one has to modulate for example $I_\mathrm{inj}$ by reversing the polarity of the charge current ($I_\mathrm{inj}=\pm I$) and applying $I_\mathrm{inj}=0$ to the injector and record $V_\mathrm{det}$ with a constant $I_\mathrm{mod}$ in these three different states. By correctly adding or subtracting these voltage values it is then possible to separate the voltage contributions from injector and modulator at the detector. We can then define the SHE induced magnon transport contribution $V_\mathrm{det}^\mathrm{inj,SHE}$ arriving from the injector at the detector
\begin{equation}
V_\mathrm{det}^\mathrm{inj,SHE}=\frac{1}{2}\left[V_\mathrm{det}\left(+I,I_\mathrm{mod}\right)-V_\mathrm{det}\left(-I,I_\mathrm{mod}\right)\right]\,.
\label{eq:dcSHE_magnon}
\end{equation}
In similar fashion, one finds the thermal contribution arriving at the detector
\begin{equation}
V_\mathrm{det}^\mathrm{inj,therm}=\frac{1}{2}\left[V_\mathrm{det}\left(+I,I_\mathrm{mod}\right)+V_\mathrm{det}\left(-I,I_\mathrm{mod}\right)-2 V_\mathrm{det}\left(0,I_\mathrm{mod}\right)\right]\,.
\label{eq:dcthermal_magnon}
\end{equation}
In this way, both injection mechanisms for magnon transport from injector to detector (SHE and Joule heating) can be separately investigated with this method.

The other experimental approach applies an ac charge current (sinusoidal with a frequency of a few Hz) to the injector $I_\mathrm{inj}(t)=I\sin(\omega t)$  and a dc charge current to the modulator. Then the contributions from injector and modulator to $V_\mathrm{det}$ can be separated by utilizing a lock-in detection as contributions from the modulator manifest as a dc signal. Moreover, as the SHE contributions scale linearly with $I_\mathrm{inj}$ and Joule heating contributions quadratic, we can detect the SHE contribution in the first harmonic signal and the thermal injection in the second harmonic signal of the lock-in detector~\cite{Cornelissen2015,Cornelissen2018}.

Both techniques allow to distinguish between SHE and Joule heating injected magnons and thus enable the investigation of magnon transport modulation in the three strip geometry. A more detailed quantitative comparison of these two measurement techniques is presented in Ref.~ \cite{Gueckelhorn_2020}. From this detailed analysis we find that for small $I_\mathrm{mod}$ both techniques provide quantitative identical results (within the noise floor of the experimental setup). A deviation is obtained if $I_\mathrm{mod}$ is increased above a threshold current, which we identify as the regime of compensation of magnetic damping in the MOI, as discussed in the following.

\subsection{Charge current induced compensation of magnon damping}
\begin{figure}
  \includegraphics[width=85mm]{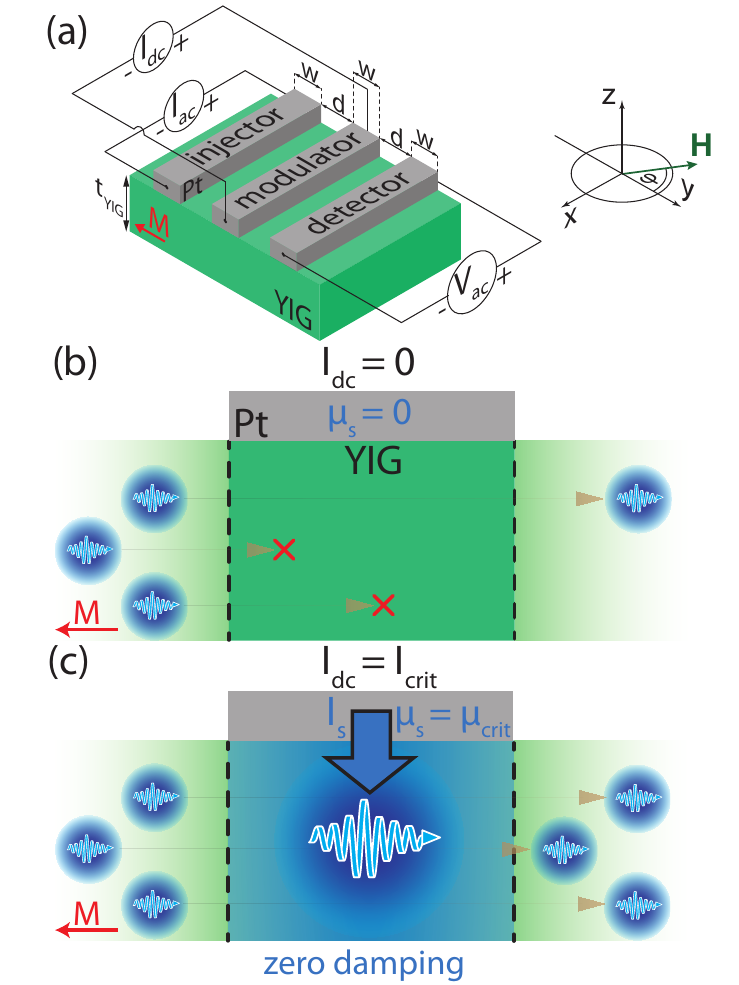}
  \caption{Magnon transport at damping compensation. (a) Depiction of the device structure and measurement scheme used in the experiment. (b) For no applied charge current to the modulator, the magnon transport between injector and detector is governed by the intrinsic relaxation rates of the YIG film. (c) At large enough charge currents $I_\mathrm{dc}\geq I_\mathrm{crit}$, the intrinsic magnon relaxation rates are compensated by the SHE-based magnon injection from the modulator (zero damping state). In this regime magnon transport underneath the modulator is greatly enhanced, also affecting the magnon transport between injector and detector. Adapted with permission.\textsuperscript{[Ref.~\cite{wimmer_2019}]} 2019, American Physical Society.}
  \label{fig:MagnonDampinScheme}
\end{figure}

The experiments discussed in this section have been conducted on YIG/Pt heterostructures and are published in Ref.~\cite{wimmer_2019}. To achieve magnetic damping compensation via SHE spin current injection requires thin YIG films with a thickness of $\approx 10\;\mathrm{nm}$. In principle even thinner YIG films would be beneficial, but magnetic damping in such YIG films also considerably increases, compensating the gain obtained by a reduction in magnetic volume. An illustration of the device geometry and the electrical measurement scheme is shown in Fig.~\ref{fig:MagnonDampinScheme}(a). The device discussed in the following consists of the $500\;\mathrm{nm}$ wide injector, modulator and detector strips, with a edge-to-edge separation of $400\;\mathrm{nm}$ between these strips. In these measurements, we apply an ac charge current to the injector and a dc charge current $I_\mathrm{dc}$ to the modulator and detect the magnon transport signal $V_\mathrm{ac}$ arriving at the detector from the injector via lock-in detection at the first harmonic.

In Fig.~\ref{fig:MagnonDampinScheme}(b) the situation underneath the modulator and the magnon transport between injector and detector for $I_\mathrm{dc}=0$ is illustrated. Due to the finite magnon relaxation rate of the YIG film, not all magnons generated at the injector reach the detector. However, as discussed above upon application of a finite $I_\mathrm{dc}$, we inject via the SHE magnons into the system. We thus provide an additional interfacial injection rate, which scales linearly with the applied dc charge current. At large enough $I_\mathrm{dc} \geq I_\mathrm{crit}$ it is possible to compensate the intrinsic magnon decay rate such that the transport of magnons between injector and detector is significantly enhanced and we establish a zero-effective damping state underneath the modulator strip (Fig.~\ref{fig:MagnonDampinScheme}(c)).

\begin{figure}
 \centering
 \includegraphics[width=85mm]{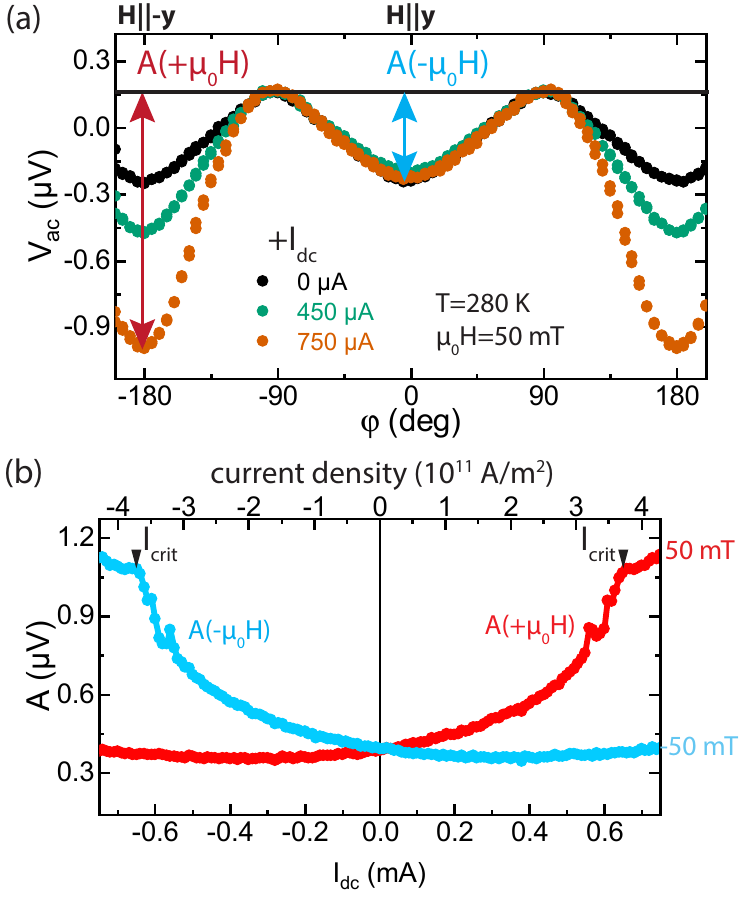}\\
 \caption[Magnon transport in zero effective damping state]{(a) Angle-dependent first harmonic signal at the detector for different $I_\mathrm{dc}$ applied to the modulator. With increasing positive $I_\mathrm{dc}$ the depth of the minima at $\pm 180 \degree$ increases, while the minimum at $0\degree$ remains unaffected. The red and blue arrows indicate the extracted amplitude $A$ for positive and negative fields, respectively. (b) Modulator current dependence of $A$ for positive and negative applied external magnetic field. For $+50\;\mathrm{mT}$ applied to the sample, $A$ increases for $I_\mathrm{dc}$ until at large $I_\mathrm{dc}$ the slope drastically changes. For negative magnetic field, the increase occurs for $I_\mathrm{dc}<0$, consistent with SHE based magnon injection from the modulator. The zero effective damping threshold $I_\mathrm{crit}$ is indicated by black triangles. Reproduced with permission.\textsuperscript{[Ref.~\cite{wimmer_2019}]} 2019, American Physical Society.}
  \label{figure:MagnonZeroDamping}
\end{figure}

In the experiment the effect of the modulator current onto magnon transport is evaluated in different measurement schemes. In Fig.~\ref{figure:MagnonZeroDamping}(a) we show the results obtained for angle-dependent measurements of the in-plane field orientation conducted at different $I_\mathrm{dc}$ (definition of $\varphi$ see Fig.~\ref{fig:MagnonDampinScheme}(a)). For $I_\mathrm{dc}=0$, we observe a $\cos^2$ angle-dependence consistent with all-electrical magnon transport experiments. If the external magnetic field and thus the magnetization of the YIG layer is oriented perpendicular to the Pt strips ($\mathbf{s}\parallel\pm \mathbf{m}$, $\varphi=\pm180\degree,0\degree$), magnon transport driven via the SHE is possible between injector and detector. If the external magnetic field is aligned along the Pt strips ($\varphi=\pm90\degree$) magnon transport between injector and detector via the SHE is suppressed and we only observe a finite offset signal. Upon application of a finite $I_\mathrm{dc}>0$, the magnon transport signal at $\varphi=\pm180\degree$ increases, while it remains nearly unaffected at $\varphi=0\degree$. This observation can be explained by the combined action of SHE and Joule heating magnon injection from the Pt modulator. At $\varphi=\pm180\degree$ and $I_\mathrm{dc}>0$ SHE and Joule heating cause a magnon accumulation underneath the modulator, which counteracts the intrinsic magnon decay rate in the YIG and thus magnon transport between injector and detector is enhanced. For $\varphi=0\degree$ and $I_\mathrm{dc}>0$, the SHE results in an increase of the magnon decay rate underneath the modulator, which is nearly compensated by the magnon injection via Joule heating. In this configuration the magnon transport between injector and detector is only weakly affected. This observation is different to the first magnon transport modulation experiments by Cornelissen \textit{et al.}~\cite{Cornelissen2018}. In those experiments an increase in amplitude for both orientations ($\varphi=\pm180\degree$,$\varphi=0\degree$) is observed. Since Cornelissen \textit{et al.} utilized much thicker YIG films for their initial experiments, the SHE contribution to the magnon transport modulation is rather weak and the thermal contribution from Joule heating in the modulator is the dominating contribution. This difference highlights the importance to use thin MOIs layers to enhance the effects of SHE based spin current injection in the experiment.

To better quantify the changes associated with $I_\mathrm{dc}$, one can extract the magnon transport amplitude $A(I_\mathrm{dc})$ for positive and negative external magnetic fields as indicated in Fig.~\ref{figure:MagnonZeroDamping}(a), the results of this procedure are shown in Fig.~\ref{figure:MagnonZeroDamping}(b). For a positive external magnetic field, and low modulator currents $A$ exhibits a linear and quadratic dependence on $I_\mathrm{dc}$, consistent with the combined contributions of SHE (linear current dependence) and Joule heating (quadratic current dependence) to the magnon relaxation rate. However, at large $I_\mathrm{dc}$ $A(I_\mathrm{dc})$ deviates from this expected behaviour and a drastic change in slope happens at a critical current value $I_\mathrm{crit}$. For negative external magnetic fields a similar drastic change in slope is found at large negative $I_\mathrm{dc}$. This field polarity dependence highlights the importance of the SHE-based contribution to the magnon relaxation rate.

The region of compensation of intrinsic magnon damping rate is associated with the manifestation of auto-oscillation of the magnetic order parameter as described in the model of spin Hall nano-oscillators (SHNOs)~\cite{Demidov2012,Demidov2014,Hamadeh2014,Collet2016,evelt_spin_2018,Evelt2018PMA,Ulrichs2020}. This describes a coherent oscillation of magnetization that is dynamically sustained via spin-transfer torque at the NM/FMI interface. The theory for SHNOs utilizes a classical Landau-Lifshitz-Gilbert (LLG) model based on a macrospin and rate equations for a FMI/NM heterostructure. For the magnon relaxation rate $\Gamma_\mathrm{mr}$ of the lowest energy mode ($k=0$), we find for an in-plane magnetized film~\cite{Stiles2006}
\begin{equation}
\Gamma_\mathrm{mr}^\mathrm{ip} = 
\left(\alpha_\mathrm{G} + \alpha_\mathrm{sp}\right)\gamma\mu_0\left(H + \frac{M_\mathrm{s}}{2}\right)\;.
\label{eq:damprate}
\end{equation}
Here, $\alpha_\mathrm{G}$ is the Gilbert damping, $\alpha_\mathrm{sp}$ the Gilbert damping induced by spin pumping due to the adjacent NM, $\gamma$ the gyromagnetic ratio, $H$ the external field and $M_\mathrm{s}$ the saturation magnetization. This relaxation rate is identical to the frequency linewidth of the $k=0$ ferromagnetic resonance mode of the FMI. The interface injection rate via SHE and interfacial spin-transfer torque determined via the spin mixing conductance is given by~\cite{Collet2016}
\begin{equation}
\Gamma_\mathrm{ST} = \frac{\hbar}{2e}\frac{\gamma}{M_\mathrm{s} t_\mathrm{FMI}}\cdot T \cdot \thetaSH I_\mathrm{dc}\;,
\label{eq:torquerate}
\end{equation}
with $t_\mathrm{FMI}$ the film thickness of the YIG layer, $T$ denotes the interface transparency for spin currents as put forward in Ref.~\cite{Zhang2015} and is given by
\begin{equation}
T = \frac{\tilde{g}_r^{\uparrow\downarrow} \tanh(\eta)}{\tilde{g}_r^{\uparrow\downarrow} + \frac{h}{2e^2}\frac{\sigma_\mathrm{NM}}{\lspinNM}}\;,
\label{eq:transparency}
\end{equation}
where $\eta = \frac{t_\mathrm{NM}}{2 \lspinNM}$ with the spin diffusion length $\lspinNM$ of the NM, the thickness $t_\mathrm{NM}$ of the NM and the electrical conductivity $\sigma_\mathrm{NM}$ of the NM assuming $t_\mathrm{NM}\gg\lspinNM$. When the condition $\Gamma_\mathrm{mr}^\mathrm{ip} = \Gamma_{\mathrm{ST}}$ is met, a coherent precession of the magnetization with zero effective damping is present~\cite{Stiles2006}. In other words, thermal fluctuations and inherent magnon interaction lead to coherent self-oscillations of the macrospin in absence of damping. As evident from these equations, it is crucial to reduce the thickness of the FMI, while keeping its Gilbert damping $\alpha_\mathrm{G}$ as low as possible, to achieve low threshold currents. In addition, sizeable SHE expressed via $\thetaSH$ is also necessary to lower the threshold current in combination with a transparent NM/FMI interface for pure spin currents. Given these prerequisites, first experiments in this direction have been conducted with YIG as the FMI and Pt as the NM~\cite{Collet2016,Evelt2018PMA,wimmer_2019}. In our all-electrical magnon transport experiments, as detailed in Ref.~\cite{wimmer_2019}, it is crucial to further account for contributions from inhomogenous line broadening, which allows to quantitatively describe the magnetic field dependence of $I_\mathrm{crit}$. 

Most importantly, the results in Ref.~\cite{wimmer_2019} show that it is indeed possible to achieve compensation of intrinsic magnon damping in thin FMI layers via a dc charge current in the three strip magnon transistor geometry. In this way, any losses underneath the modulator are compensated giving access to a new diffusive magnon transport regime without magnon spin relaxation. This damping compensation effect via a dc charge current may allow all-electrical experimental studies of magnon transport in the limit of spin superfluidity and magnon Bose-Einstein condensates~\cite{Demokritov2006,Demidov2007,Demidov2008,Bender2012,Bender_2014,Shen2019}  as discussed in recent theoretical publications~\cite{Flebus2016,nakata_spin_2017,Rckriegel2017,Iacocca2017,Troncoso2019,Takei2019,Sonin2019,Iacocca2019a,Iacocca2019,Evers2020,Sonin2020}. For example Ref.~\cite{Takei2019} provides analytical expressions for the evolution of the magnon transport close to the point of damping compensation. Another aspect not covered in this brief discussion are contributions from localized magnon modes, which in general exhibit lower threshold values for achieving damping compensation and may also affect the magnon transport~\cite{Ulrichs2020}. However, a full understanding of all relevant processes and their influence on magnon transport has still to be obtained. New results shine some more light onto the role of Joule heating in these experiments~\cite{liu2020electrically}. In addition, similar effects have been observed using a ferromagnetic metal as the modulator strip~\cite{Santos2021}. Hopefully more experiments and a better theoretical understanding of the non-equilibrium and chaotic processes at work (especially in the regime beyond magnetic damping compensation) allow to further explore the regime of spin superfluidity and magnon Bose-Einstein condensates via all-electrical magnon transport experiments in MOIs at the nanoscale.

While we focused in this section on the effects in FMIs it is also important to discuss if such magnon transport modulation experiments are possible in AFIs. Here, things get more complex and interesting as we find two magnon modes with opposite spin in these AFI systems. We here follow the theoretical calculations by Shen~\cite{Shen2019}. As discussed in this section, the SHE modifies the spin relaxation rate in the MOI. For an AFI the $\alpha$- and $\beta$-magnons react oppositely to the SHE induced damping contribution. Thus, for one current polarity and fixed orientation of $\mathbf{n}$, one mode exhibits enhanced damping, while the other features reduced damping. If we reverse the current polarity the situation is inverted for these two modes. This is in contrast to the discussion above for FMIs, where we only had to consider one magnon mode. In the framework of the proposed two magnon spin channel (and assuming distances shorter than $\lspin$), we can employ a parallel resistor model to describe the changes associated with the modulator current. This will lead to much more complex evolution of the magnon transport between injector and detector as a function of $I_\mathrm{mod}$. Especially, one needs to account for the effects of Joule heating in these devices. Moreover, Shen~\cite{Shen2019} predicts similarly as in the case for FMI~\cite{Bender2012,Bender_2014} a critical current to reach the formation of a magnon Bose-Einstein condensate via dc charge current pumping for one of the magon modes. As discussed in Ref.~\cite{wimmer_2019}, the criteria for the formation of magnon Bose-Einstein condensates can be also discussed in the framework of damping compensation and auto-oscillations. It is thus worthwhile to mention that the effect of damping compensation in AFIs has been discussed theoretically in the context of spin Hall nano-oscillators~\cite{Cheng2016,khymyn_antiferromagnetic_2017,Sulymenko2017,Lee2019} and spin-transfer torque effects~\cite{Gomonay2010,Gomonay2014}. These works suggest that it is indeed possible to generate auto-oscillations in the case of easy-axis as well as easy-plane magnetic anisotropy in antiferromagnets. Yet, depending on the magnetic anisotropy a certain alignment between $\mathbf{s}$ and $\mathbf{n}$ is required to achieve stable oscillations, which might require special experimental geometries and/or magnetically ordered metals as the modulator~\cite{Lee2019}. However, there are presently no experimental studies available that venture in this direction, but this would be an interesting next step to take for AFIs.

\section{Magnon Pseudospin dynamics in antiferromagnetic insulators}
\label{sec:magnonPseudo}
\begin{figure}
  \includegraphics[width=85mm]{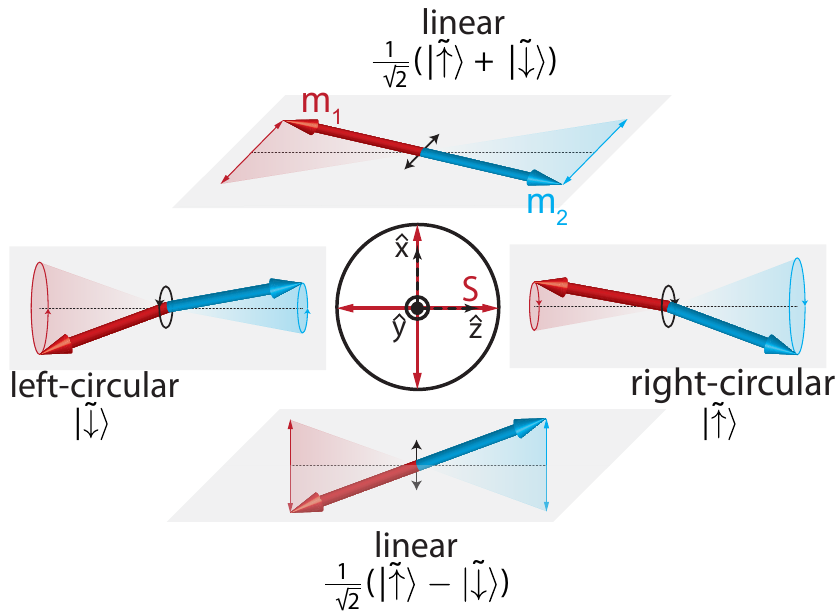}
  \caption{Concept of Pseudospin in antiferromagnetic insulators. Within the concept of pseudospin, the left- and right-circular polarized magnon modes represent the spin-down and spin-up states of the pseudospin,i.e.~the pseudospin is collinear with the z-axis. The linear polarized magnon modes are superpositions of the left- and right-circular magnon modes, representing zero-spin excitations. The dynamical coupling between these different magnon modes can be described via a rotation of the pseudospin around the y-axis. The magnonic spin investigated in all-electrical magnon transport experiments is determined by the z-component of the pseudospin. Reproduced with permission.\textsuperscript{[Ref.~\cite{wimmer_observation_2020}]} 2020, American Physical Society.}
  \label{fig:MagnonPseudospin}
\end{figure}

In the previous sections, we assumed that in the case of AFIs, we have two magnon modes carrying opposite spin that both contribute to the spin transport via magnons. It is also possible to combine these two different magnon modes and form linearly polarized magnon modes carrying zero spin. We can describe this pair of magnons and their superositions via a pseudospin~\cite{Cheng2016,Daniels2018,Kawano2019,Shen2020,kamra_antiferromagnetic_2020,wimmer_observation_2020} quite similar like the spin of an electron (see Fig.~\ref{fig:MagnonPseudospin}). In the description employed here, the detected spin in our all-electrical magnon transport experiments corresponds to the z-component of the pseudospin. In our recent publications, we showed that this pseudospin property is experimentally accessible in AFIs leading to the observation of the magnon Hanle effect~\cite{kamra_antiferromagnetic_2020,wimmer_observation_2020}. These results how that we obtain coherent control over magnon spin transport confirmed by further recent experiments~\cite{Ross2020}.

\begin{figure}
  \includegraphics[width=85mm]{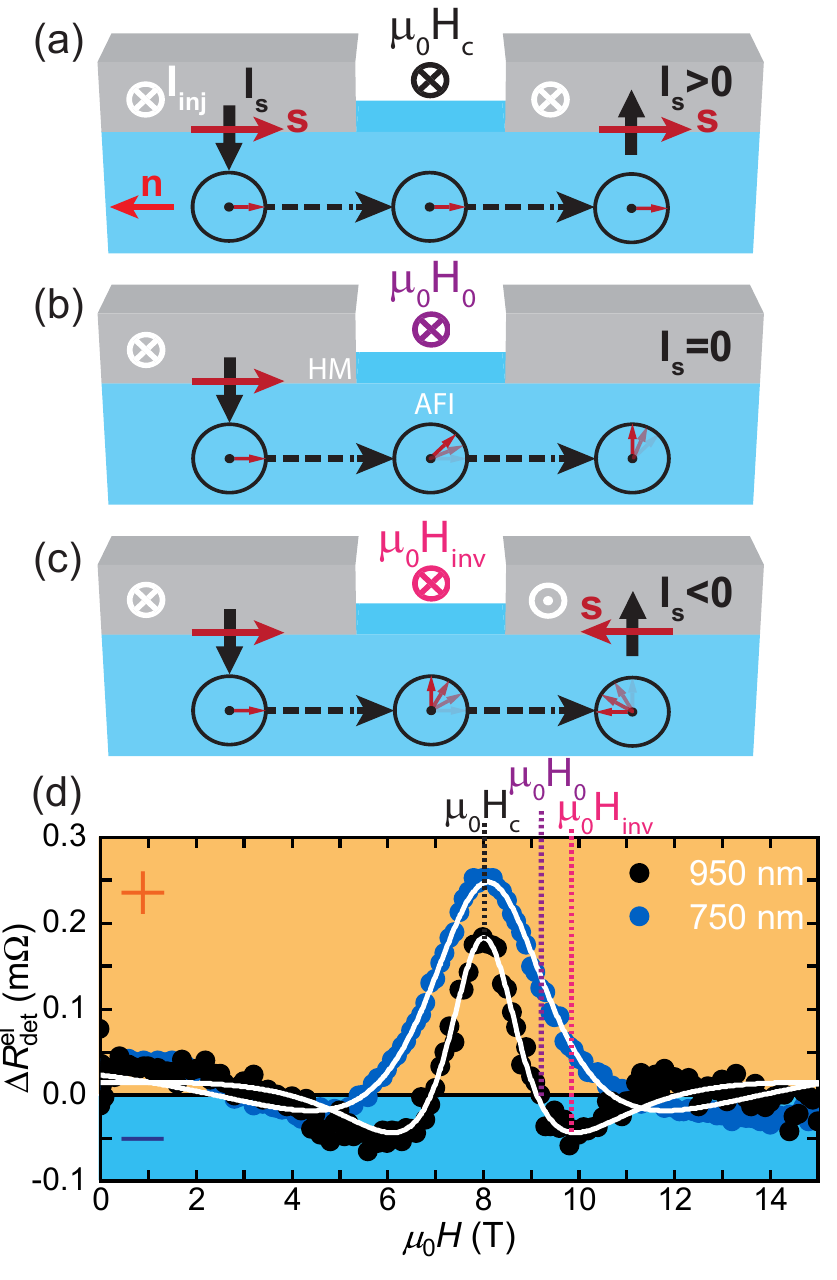}
  \caption{Illustration of the Magnon Hanle effect. A pseudospin density polarized in $z$-direction is injected via the NM strip and propagates in the AFI to the NM detector strip. The etxernal magnetic field allows to control the precession frequency of the pseudospin in the $x-z$-plane. This results in positive (a), zero (b) and negative (c) spin signals arriving at the detector. (d) Measured spin signal for two injector-detector distances (black and blue circles) as a function of the external magnetic field at $T=200\;\mathrm{K}$. White lines represent fits to the obtained data via Eq.(\ref{eq:fit}). Reproduced with permission.\textsuperscript{[Ref.~\cite{wimmer_observation_2020}]} 2020, American Physical Society.}
  \label{fig:MagnonHanle}
\end{figure}

In more detail, we inject via the SHE a spin current into the AFI, which generates a magnon pseudospin density $\mathbf{\mathcal{S}}$ oriented along $z$ underneath the injector (see Fig.~\ref{fig:MagnonHanle} (a)-(c)). In the $\alpha$-Fe$_2$O$_3$ thin films we investigated, the magnetic easy-plane anisotropy and the Dzyaloshinskii-Moriya interaction (DMI) cause a coherent coupling between the spin-up and spin-down magnons, which leads to a precession of the pseuodspin in the $x$-$z$ plane. The magnetic anisotropy and the combination of DMI field and canting-induced net magnetization determine the pseudospin precession frequency. The canting-induced net magnetization depends on the external magnetic field, such that we can tune the precession frequency via the application of an external magnetic field. Three situations are illustrated in Fig.~\ref{fig:MagnonHanle}(a)-(c). At the compensation field $H_\mathrm{c}$ the contributions from anisotropy and DMI cancel each other, such that no precession occurs and the pseudospin just propagates through the AFI. In this situation the same spin orientation injected via SHE is detected at the NM detector strip, resulting in a positive spin signal. For $H_0$, the finite pseuodspin precession causes the pseudospin of the magnons arriving at the detector strip to be oriented perpendicular to the $z$-direction, corresponding to a spin zero state. Thus, no spin signal is detected. At $H_\mathrm{inv}$, the magnon pseudospin has reversed direction while propagating from injector to detector, which results in a negative spin signal at the detector.

As discussed in more detail in Ref.~\cite{kamra_antiferromagnetic_2020}, the diffusive transport and dynamics of $\mathbf{\mathcal{S}}$ in the case of an easy-plane AFI with finite DMI is described via
\begin{equation}
    \frac{\partial \mathbf{\mathcal{S}}}{\partial t}  =  D_\mathrm{mag,s} \nabla^2 \mathbf{\mathcal{S}} - \frac{\mathbf{\mathcal{S}}}{\tau_\mathrm{s}} + \mathbf{\mathcal{S}} \times \Omega \, \hat{\mathbf{y}}\;,
	\label{eq:pseudospin_diffusion}
\end{equation}
resembling the equations for electronic spin transport in an magnetic field. In comparison to the derived magnon spin diffusion equation Eq.(\ref{eq:spindiff}), we here introduce an additional pseudofield contribution $\Omega$, characterizing the pseudospin precession. In our model, we find a linear dependence of $\Omega$ on the external magnetic field $\mu_0H$~\cite{kamra_antiferromagnetic_2020,wimmer_observation_2020}. We solve Eq.(\ref{eq:pseudospin_diffusion}) by accounting for the injection of a $z$-polarized pseudospin current density $j_\mathrm{s0}$ at the injector at $z=0$. One then obtains for the $z$-component of the pseudospin density (the spin signal at the detector) at a distance $d$ away from the injector
\begin{equation}
    	\mathbf{\mathcal{S}}_\mathrm{z}(d) = \frac{j_\mathrm{s0}\lambda_\mathrm{sf}}{D_\mathrm{mag,s}(a^2+b^2)}\mathrm{e}^{-\frac{a d}{\lambda_\mathrm{sf}}}\left(a\cos{\frac{b d}{\lambda_\mathrm{sf}}}-b\sin{\frac{b d}{\lambda_\mathrm{{sf}}}}\right)\;,
	\label{eq:fit}
\end{equation}
where $a \equiv \sqrt{(1+\sqrt{1+\Omega^2\tau_\mathrm{s}^2})/2}$, $b \equiv\sqrt{(-1+\sqrt{1+\Omega^2\tau_\mathrm{s}^2})/2}$, and $\lambda_\mathrm{sf} = \sqrt{D_\mathrm{mag,s} \tau_s}$ is the magnon spin diffusion length. The equation exhibits an oscillating part (as a function of distance $d$ and external magnetic field) with an exponential envelope function. The description is very similar to the electronic Hanle effect~\cite{Jedema2002,Fabian2007} now the pseudofield $\Omega$ is responsible for the pseudospin precession as compared to the external magnetic field applied perpendicular to the spin polarization in case of the electronic Hanle effect. In both situations, we find a maximum spin signal for no (pseudo)spin precession ($\Omega=0$). However, for the electronic Hanle effect this maximum is positioned at zero external applied magnetic field, while for the magnon Hanle effect one finds the maximum at a finite external magnetic field corresponding to the compensation field $H_\mathrm{c}$. In addition, the dependence of $\Omega$ on the external magnetic field depends on the properties of the AFI and it is possible, that not only a linear dependence can be observed, but also higher order contributions could play an important role.

In our first observation of the magnon Hanle effect~\cite{wimmer_observation_2020}, we utilized a $15\;\mathrm{nm}$ thin $\alpha$-Fe$_2$O$_3$ layer and $4\;\mathrm{nm}$ Pt for the NM strips on top. All-electrical magnon transport experiments have been conducted in the two strip geometry with Pt injector and detector utilizing the current reversal technique to analyze the SHE induced magnon transport signal at the detector. Utilizing such a thin AFI layer provides us with the advantage, that the magnon diffusion is predominantly 1d, such that the assumptions of our employed model are easily transferable to the experiment. In addition, the $\alpha$-Fe$_2$O$_3$ remains in the easy-plane magnetic anisotropy at all investigated temperatures, providing a large parameter space to conduct experiments. Since in the AFI the N\'eel-vector $\mathbf{n}$ orients perpendicular to the external magnetic field~\cite{hoogeboom_negative_2017,ji_spin_2017,hou_tunable_2017,manchon_spin_2017,fischer_spin_2018,Klaui2018,Fischer2020,Geprgs2020,Han2020}, SHE induced magnon transport is possible if the external magnetic field is aligned in-plane along the NM strips. To analyze the magnetic field dependence of the magnon transport signal at the detector, measurements in both in-plane field geometries, parallel and perpendicular to the strips, are necessary to remove additional weak background signals.

The results of these measurements for $d=750\;\mathrm{nm}$ (blue circles) and $d=950\;\mathrm{nm}$ (black circles) are shown in Fig.~\ref{fig:MagnonHanle}(d). Most strikingly, at $\mu_0 H= 8\;\mathrm{T}$ we find a maximum positive signal at the detector for both distances $d$. This corresponds to the compensation field $H_\mathrm{c}$, at which $\Omega=0$ and no pseudospin precession occurs. $H_\mathrm{c}$ only depends on the intrinsic properties of the AFI and is independent of the distance $d$ between injector and detector in agreement with our experiments. Upon further increasing the field above $H_\mathrm{c}$, the signal at the detector first decreases to $0$ (corresponding to $H_0$), then becomes negative with a minimum at $H_\mathrm{inv}$. At even higher fields, the signal at the detector becomes again positive. A similar magnetic field dependence is observed for decreasing fields from $H_\mathrm{c}$. As expected, the extracted field-dependent magnon transport signals are comparable to the electronic Hanle signals observed in non-local spin valve structures~\cite{Jedema2002}. Our pseudospin model nicely explains the data as indicated by the white lines, which are a fit to each data set via Eq.(\ref{eq:fit}). We can extract from these measurements the magnon diffusion length and obtain values of $\approx 300\;\mathrm{nm}$. In Ref.~\cite{wimmer_observation_2020}, we conducted further control experiments. Via angle-dependent field orientation experiments, we confirm the sign change in the magnon transport signal and the N\'eel vector origin of the magnon transport signal. Further temperature-dependent measurements show that the compensation field $H_\mathrm{c}$ in our $\alpha$-Fe$_2$O$_3$ layer saturates for low temperatures and increases with increasing temperature. The extracted magnon spin diffusion length is only weakly temperature dependent. Most likely due to the fact that magnon scattering on $180\degree$ domain walls in the $\alpha$-Fe$_2$O$_3$ dominates the magnon spin diffusion length~\cite{Klaui2020}.

We are still at the early stages of obtaining a deeper understanding of the pseudospin and pseudofield in antiferromagnetic insulators. Especially, a full quantitative understanding of the microscopic origins of the pseudofield and its dependence on experimentally accessible control parameters has to be obtained in the future. Yet, our pseudospin model~\cite{kamra_antiferromagnetic_2020} provides an elegant way to describe the complex coupling of the magnon eigenstates in an easy-plane antiferromagnet and the transport of spin information within such a system. In this section, we only covered the oscillating component originating from the pseuodfield, but additional components of the pseudofield can also cause changes to an additional spin transport signal, which in our first experiments was assumed to be constant with respect to the applied external magnetic field. Our results show that indeed aspects of electronic spin transport can be mapped onto magnon spin transport in antiferromagnets and results in experimentally accessible effects. Thus, magnon transport in antiferromagnetic insulators is a promising playground to realize magnonic analogues of electron spin transport~\cite{mook_taking_2018,Daniels2018,Shen2019,kamra_antiferromagnetic_2020}. A number of theoretical proposals may now be realized in this class of materials ranging from emerging spin-orbit coupling over topological surface states to nonabelian computing schemes~\cite{nakata_magnonic_2017,Daniels2018,Kawano2019,Kawano2019B,li_magnon_2020-1,Mook2020,malki_topological_2019,malki_topological_2020,nayga_magnon_2019,dos_santos_modeling_2020,Evers2020,Shen2020,liu_dipolar_2020,li_magnon_2020,zhang_magnon_2020}.

\section{Conclusion and Outlook}

Since the first theoretical prediction and experimental demonstration of all-electrical magnon transport in NM/MOI heterostructures~\cite{ZhangMMR1,ZhangMMR2,Cornelissen2015}, experiments in this regard were carried out not only in ferrimagnetically ordered insulators~\cite{Goennenwein2015,Cornelissen2016_temperaturedependence,Cornelissen2016_fielddependence,shan_influence_2016,Cornelissen2016,Shan2017_NFO,Das2017,Shan_2018_NFO,Cornelissen2018,Liu2018,Thiery_2018,Shen2019,wimmer_AHE_2019,wimmer_2019,Avci2020,Troncoso2020}, but also in antiferromagnetic insulators~\cite{Klaui2018,Klaui2020,Han2020,wimmer_observation_2020,Lebrun2020,Ross2020}. In this way, we obtained access to studying magnon transport, corresponding to GHz or even THz frequencies, utilizing magnetotransport techniques and dc charge currents. These experiments pave the way towards investigating diffusive magnon spin transport down to the nanoscale. Moreover, with the improvement in interface properties and materials the investigation of effects beyond linear response with respect to the applied charge current, \ie~SHE-based spin current injection and Joule heating induced temperature gradients, are accessible and provide means to control the magnon transport in MOIs. Interestingly, in very thin MOIs, it is possible to locally compensate the damping of magnetization dynamics via charge current bias and drive auto-oscillations/magnetization dynamics in the MOI~\cite{Demidov2012,Demidov2014,Hamadeh2014,Collet2016,evelt_spin_2018,Evelt2018PMA}. In this way, dc charge currents generate high frequency excitations in the magnetic lattice and provide even access to THz magnonics in antiferromagnetic insulators. Moreover, in the regime of damping-compensation, magnon Josephson physics is within reach in these type of experiments~\cite{Demokritov2006,Demidov2007,nakata_spin_2017,Bozhko2019}. In antiferromagnetic insulators, the intrinsic two magnon modes of the magnetic system can be considered as the bosonic analogue of an electron spin-1/2 system~\cite{mook_taking_2018,Shen2019,kamra_antiferromagnetic_2020}. This analogy has now been verified in all-electrical magnon transport experiments~\cite{wimmer_observation_2020} and promises new physics closely related to electronic spin transport in antiferromagnetic magnonics. 

All-electrical magnon transport experiments heavily rely on efficient means to generate and detect pure spin currents via charge currents. Thus, from a materials point of view it is imperative to further develop not only more efficient means for charge-to-spin current conversion, but also explore novel functionalities provided by new materials~\cite{Han2018}. There are several directions to further explore. For example, charge-to-spin current conversion in magnetic conductors, exhibiting ferromagnetic or antiferromagnetic order, has received a lot of attention in the last years and many aspects are yet to be explored. These materials offer the unique tunability of the flow direction and spin orientation of the generated pure spin current via the orientation of the magnetic order parameter providing novel design parameters for spin-transfer torque devices~\cite{Taniguchi2015,Kimata2019,Zhang2019_Weyl,Davidson2020,Mook2020}. Moreover, in these magnetically ordered conductors, excitations of the magnetic lattice can interact with the conduction electrons in the system, providing additional and unexplored aspects of pure spin current transport and spin-to-charge conversion in these systems. Another important direction are 2d materials for charge-to-spin current conversion~\cite{Safeer2019,Ghiasi2019,Vaz2019,Rossi_2019,Liu_2020}, where nowadays with the possibility to tune the properties of the materials via stacking 2d materials on top of each other provides a vast parameter space to obtain efficient pure spin current generation in such 2d systems. In addition, topological insulators with their spin polarized edge state provide novel functionalities in charge-to-spin current conversion processes~\cite{Wang_2016,Zhang_2016,Han2018,Wang_2019}.

Developing new MOIs to unlock new areas of magnon transport is another important aspect for material oriented research. One of the major challenges is to establish MOIs with long magnon spin lifetimes to facilitate experiments on lengthscales accessible via nanolithography processes. At present a large quantity of experiments utilized YIG layers for the experiment as this garnet material exhibits long magnon spin lifetimes. The excellent properties of YIG thin films is strongly linked to the availability of gadolinium gallium garnet as a lattice matched substrate, exhibiting an extremely low lattice mismatch with YIG. Similar advances may also be possible in different classes of materials, if suitable substrates are available. One example in this regard are nickel ferrite (NFO) thin films, where magnesium gallate substrates significantly reduce the Gilbert damping and thus increase the magnon spin lifetime~\cite{singh_bulk_2017}. In all-electrical magnon transport experiments~\cite{Shan_2018_NFO} NFO grown on such lattice matched substrates exhibits similar properties than YIG thin films. Similar enhancement in magnontransport properties may be found in other materials, especially further development in the growth of AFIs on lattice matched substrates has the potential to significantly enhance magnon transport properties in thin films and match their bulk counterparts. In very thin MOI layers, we gain the chance to investigate the effect of reduced dimensionality on magnon transport. 2d van der Waals magnets provide also access to the field of 2d MOIs~\cite{Burch_2018,Zhang_2019_2dmag}. Such 2d MOIs open up directions of research beyond magnetic order in 3d systems and act as a new system to test established theoretical concepts of magnon transport. Another important aspect, where of the last two years tremendous progress has been made is the investigation of interfacial Dzyaloshinskii–Moriya interaction(DMI) and the generation of topological magnetic textures in heterostructures based on MOIs~\cite{Ding2019,Caretta2020,Xia2020_chiral,Wang2020_chiral,Lee2020_chrialnano,Ding2020}. These results promise future progress quite similar as in metallic multilayer systems~\cite{Wiesendanger2016,Dup2016,Fert2017,Soumyanarayanan2017,Hellman2017,Maccariello2018,EverschorSitte2018,Back2020}. In this regard, also a development of theoretical concepts explaining the sometimes contradicting experimental results is desirable. Nevertheless, the engineering of interfacial DMI in MOIs opens up new avenues to venture towards functional materials for magnonics. In similar fashion, multilayer systems forming insulating, synthetic ferri- or antiferromagnets open up a promising direction to control and engineer novel magnetic phases in MOI systems.

Another important aspect is a better understanding how to engineer the NM/MOI interfaces to increase the spin transparency of the interface~\cite{Ptter2017,Fontcuberta2019,Bai2019_YIG,Liu2020_spinpump,Kohno2021}. As pointed out in this review, from the point of applications it would be highly desirable to further enhance $g$ and $\tilde{g}^{\uparrow\downarrow}$ and figure out means to control these parameters by the choice of materials and interface preparation methods.

In this review, we discussed means to manipulate magnon transport via charge currents. At present all experiments in this direction utilized YIG layers as the MOI. Even in this material system a full understanding of the non-equilibrium processes contributing to the observed signatures in magnon transport is missing. Experimental techniques providing spatial resolution and imaging magnetic excitations in MOIs can provide a deeper insight especially in the regime of damping compensation. Here, magnetic imaging via nitrogen vacancies in diamond and associated scanning probe techniques provide nanometer scale resolution and the ability to investigate locally the magnon chemical potential~\cite{Du2017,wang2020quantum}. It would be highly interesting to investigate these charge current induced changes in MOIs with different magnetic order to verify the assumed theoretical models. Clearly, a description of the dynamics in the MOI in this non-equilibrium regime is challenging as chaotic dynamics may be encountered, but it will be insightful to see if such an effective zero damping state allows the investigation of magnon Bose-Einstein condensates and related phenomena~\cite{Flebus2016,nakata_spin_2017,Rckriegel2017,Iacocca2017,Troncoso2019,Takei2019,Sonin2019,Iacocca2019a,Iacocca2019,Evers2020,Sonin2020,Tserkovnyak_2020}.

\begin{figure}
  \includegraphics[width=85mm]{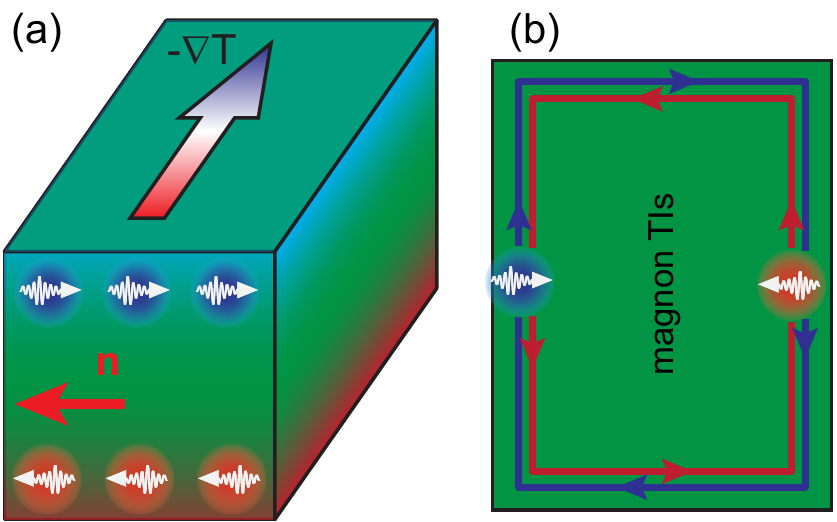}
  \caption{Emergent spin-orbit effects for magnon transport in antiferromagentic insulators. (a) Illustration of the magnon spin Nernst effect. Upon application of a thermal gradient emergent spin-orbit effects cause a magnon accumulation at the boundaries of the material. (b) Depiction of a magnon topological insulator. Emergent spin-orbit coupling phenomena induce magnon edge states whit a spin-momentum locking, i.e. magnons with different helicity move in opposite directions.}
  \label{fig:Outlook}
\end{figure}


The field of antiferromagnetic magnonics has received exceedingly more interest in the last years~\cite{jungwirth_antiferromagnetic_2016,Baltz2018}. Since the natural frequency of magnons in antiferromagnets is THz, NM/AFI heterostructures allow to realize spin Hall nano-oscillators operating at THz frequencies, enabling dc to THz conversion~\cite{Cheng2016}. An experimental realization should be possible within the next years. In antiferromagnets we find two magnon modes carrying opposite spin, similar as the two spin states in electronic spin transport. The recent observation of the magnon Hanle effect in AFIs~\cite{wimmer_observation_2020,Ross2020} showed that indeed these properties are experimentally accessible and lead to dramatic effects for the magnon transport in AFIs. Many theoretical concepts exploit the intrinsic two-level properties of magnons in antiferromagnets and the coupling to for example magnetic anisotropy or diplolar fields to construct emergent spin-orbit phenomena in analogy to electronic transport~\cite{Cheng2016_AntiFET,Cheng2016B,Mook2017,Daniels2018,Hou2019,Kawano2019,Kawano2019B,Daniels2019,Shen2020}. In terms of emergent spin-orbit coupling the magnon spin Nernst effect~\cite{Cheng2016B,Shen2020}, \ie~the generation of a transverse magnonic pure spin current by driving a magnon current via a thermal gradient, is one of the prototype examples awaiting experimental verification (see Fig.~\ref{fig:Outlook}(a)). A major challenge in the experiment will be the detection of the transverse magnon spin accumulation and careful control experiments to disentangle the magnon spin Hall effect from other effects driven by the applied thermal gradient. Moreover, several proposals to realize topological magnon insulators, where edge channels provide momentum-locked pure spin current transport, have been put forward (see Fig.~\ref{fig:Outlook}(b))~\cite{Daniels2019,Kawano2019}. If such effects can be verified in the experiment, they will provide fruitful grounds for new functionalities in magnonic applications. In regards to applications and further developments in the field, it will be crucial to determine efficient ways to tune and engineer emergent spin-orbit effects in antiferromagnetic magnonics. External control parameters may be electric fields, modulating the local electron states, as well as strain, which couples via magnetoelastic effects to the magnonic system. In regards to strain effects, the first experimental observation of magnon-phonon coupling effects in AFIs~\cite{Li_2020_magphon} provides positive evidence for the feasibility of this approach. Advances in the field require a combined effort from the side of available materials and also from spectroscopy to investigate magnon transport in antiferromagnets. In regards to spectroscopy, the proposed emergent spin-orbit effects will not only influence incoherent excitation/detection mechanism as used in dc all-electrical magnon transport experiments, but also should play a major role in coherently excited magnon transport experiments. The first step into this direction has been experimentally achieved by electrically detected spin pumping experiments in AFI/NM heterostructures~\cite{Li_2020_antispinpumping,Vaidya_2020}. Such coherent schemes will also be of importance for nonabelian computing schemes in AFIs~\cite{Daniels2018}.

Many interesting facets are pending their experimental verification and theoretical description. The combined research effort of many groups will boost this quickly developing field to the next level and lead to the realization of magnonic applications.


\medskip

\medskip
\textbf{Acknowledgements} \par 
Financial support from the Deutsche Forschungsgemeinschaft (DFG, German Research Foundation) under Germany’s Excellence Strategy -- EXC-2111 -- 390814868 and project AL2110/2-1 is gratefully acknowledged. Moreover, I would like to thank the following people for fruitful discussions, helpful input while writing this article and dedicating their resources to our results discussed here: S. Gepr\"{a}gs, S.T.B. Goennenwein, R. Gross, H. Huebl, A. Kamra, M. Opel, M. Weiler. In addition, I would like to thank B. Coester, L. Flacke, J. G\"{u}ckelhorn,  E. Karadza, L. Liensberger, S. Matsura, M. M\"{u}ller, T. Narr, K. Rubenbauer, J. Schirk, N. Vlietstra, T. Wimmer from the Walther-Mei{\ss}ner-Institut, and G.E.W. Bauer, H. Ebert, A. Gupta, M. Kl\"{a}ui, T. Kuschel, G. Reiss,  R. Schlitz, H. Ulrichs, B. van Wees and many more unsung heroes for fruitful collaborations and discussions on all-electrical magnon transport in MOIs.

\medskip

%
\bibliography{Bibliography}




\end{document}